\documentclass[traditabstract]{aa} 
\usepackage{graphicx}
\usepackage{natbib}
\usepackage{epstopdf}
\usepackage{color}
\usepackage{txfonts}
\newcommand{\kep}{{\it Kepler}}
\newcommand{\teff}{$T_{\rm eff}$}

\newcommand{\logg}{$\log g$}

\newcommand{\mlsep}{$\langle \Delta\nu \rangle$}
\newcommand{\mssep}{$\langle \delta\nu_{02} \rangle$}
\newcommand{\numax}{$\nu_{\rm max}$}
\newcommand{\feh}{[Fe/H]}
\newcommand{\vsini}{$v \sin i$}
\newcommand{\feone}{Fe-I}
\newcommand{\fetwo}{Fe-II}
\newcommand{\rad}{$R$}
\newcommand{\mass}{$M$}
\newcommand{\age}{$\tau$}
\newcommand{\rsol}{R$_{\odot}$}
\newcommand{\msol}{M$_{\odot}$}
\newcommand{\lum}{$L$}
\newcommand{\lsol}{L$_{\odot}$}
\newcommand{\kms}{kms$^{-1}$}
\newcommand{\mhz}{$\mu$Hz}
\newcommand{\vturb}{$\xi_{\rm t}$}
\newcommand{\vrad}{$v_{\rm R}$}
\newcommand{\md}{$\langle \rho \rangle$}
\newcommand{\pmm}{$\pm$}

\def\numax{$\nu_{\rm max}$}
\def\teff{$T_{\rm eff}$}
\def\logg{$\log g$}
\begin{document}
\title{Fundamental properties of five \kep\ stars using global
asteroseismic quantities and ground-based observations}

   \author{Orlagh~L.~Creevey\inst{1,2,3} 
 G\"ulnur~Do{\u g}an\inst{4,5},
Antonio~Frasca\inst{6},
Anders~Overaa~Thygesen\inst{4,7},
Sarbani~Basu\inst{8},
Jishnu~Bhattacharya\inst{9},
Katia~Biazzo\inst{10},
Isa~M.~Brand\~ao\inst{11,12},
Hans~Bruntt\inst{4},
Anwesh~Mazumdar\inst{13},
Ewa~Niemczura\inst{14},
Tushar~Shrotriya\inst{15},
S\'ergio~G.~Sousa\inst{11},
Dennis~Stello\inst{16},
Akshay~Subramaniam\inst{17},
Tiago~L.~Campante\inst{4,11,12},
Rasmus~Handberg\inst{4},
Savita~Mathur\inst{5},
Timothy~R.~Bedding\inst{16},
Rafael~A.~Garc\'ia\inst{18},
Clara~R\'egulo\inst{1,2},
David~Salabert\inst{1,2,3},
Joanna~Molenda-\.{Z}akowicz\inst{14},
Pierre-Olivier~Quirion\inst{19},
Timothy~R.~White\inst{16}
Alfio~Bonanno\inst{6},
William~J.~Chaplin\inst{20},
J{\o}rgen~Christensen-Dalsgaard\inst{4},
Jessie L. Christiansen\inst{21},
Yvonne~Elsworth\inst{20},
Michael~N.~Fanelli\inst{22},
Christoffer~Karoff\inst{4}, 
K.~Kinemuchi\inst{22},
Hans Kjeldsen\inst{4},
Ning~Gai\inst{8, 23}, 
M\'ario~J.P.F.G~Monteiro\inst{11,12}, \and
Juan~Carlos~Su\'arez\inst{24}
}

\institute{IAC
Instituto de Astrof\'{i}sica de Canarias, C/ V\'ia L\'actea s/n,
E-38200 Tenerife, Spain. \email{ocreevey@oca.eu},
\and Universidad de La Laguna, Avda. Astrof\'isico 
Francisco S\'anchez s/n, 38206 La Laguna, Tenerife, Spain.
\and  Universit\'e de Nice, Laboratoire Cassiop\'ee, CNRS UMR 6202,
Observatoire de la C\^ote d'Azur, BP 4229, 06304 Nice cedex 4, France.
\and Department of Physics and Astronomy, Aarhus University, DK-8000 Aarhus C, Denmark.
\and High Altitude Observatory, NCAR, P.O. Box 3000, Boulder, CO 80307, USA.
\and INAF, Osservatorio Astrofisico di Catania, via S. Sofia, 78, I-95123 Catania, Italy.
\and Nordic Optical Telescope, Apartado 474, E-38700 Santa Cruz de La Palma, Santa Cruz de Tenerife, Spain.
\and Department of Astronomy, Yale University, P.O. Box
208101, New Haven, CT 06520-8101, USA.
\and Dept. of Physics, Indian Institute of Technology Kanpur,
Kanpur 208016, India.
\and INAF, Osservatorio Astronomico di Capodimonte, Salita Moiariello 16, I-80131 Napoli, Italy.
\and Centro de Astrof\'isica, Universidade do Porto, Rua das Estrelas, P-4150-762 Porto, Portugal.
\and Departamento de F\'isica e Astronomia, Faculdade de Ci\^encias da Universidade do Porto,
Rua do Campo Alegre 687, P-4169-007 Porto, Portugal.
\and Homi Bhabha Centre for Science Education (TIFR), V.\ N.\ Purav
Marg, Mankhurd, Mumbai 400088, India.
\and Instytut Astronomiczny, Uniwersytet Wroc{\l}awski, ul. Kopernika 11,
51-622, Wroc{\l}aw, Poland.
\and Indian Institute of Science Education and Research Pune,
Sai Trinity Building, Garware Circle, Pashan, Pune 411021, India.
\and Sydney Institute for Astronomy (SIfA), School of Physics, University of Sydney, NSW 2006, Australia.
\and Indian Institute of Technology Madras, Chennai-600036,
India.
\and Laboratoire AIM, CEA/DSM-CNRS-Universit\'e Paris Diderot, IRFU/SAp, centre de Saclay, 91191, Gif-sur-Yvette, France.
\and Canadian Space Agency, 6767 Route de l'A\'{e}roport, Saint-Hubert, QC, J3Y 8Y9 Canada.
\and School of Physics and Astronomy, University of
Birmingham, Edgbaston, Birmingham B15 2TT, UK.
\and SETI Institute/NASA Ames Research Center, Moffett Field, CA 94035, USA.
\and Bay Area Environmental Research Inst./NASA Ames Reseach Center, Moffett Field, CA 94035, USA.
\and Department of Physics, Dezhou University, Dezhou 253023, P. R. China.
\and Instituto de Astrof\'{\i}sica de Andaluc\'{\i}a (CSIC). Glorieta de la Astronom\'{\i}a S/N. CP3004, Granada, Spain.
             }

   \date{Received September 15, 1996; accepted March 16, 1997}

 
  \abstract
 {We present an asteroseismic study of the solar-like stars KIC~11395018,
KIC~10273246, KIC~10920273, KIC~10339342, and KIC~11234888 using 
short-cadence time series of more than eight months 
from the \kep\ satellite.  
For four of these stars, we derive atmospheric parameters from spectra
acquired with the Nordic Optical Telescope. 
The global seismic quantities (average large frequency separation and frequency of maximum power),
combined with the atmospheric parameters, yield the mean density and 
surface gravity with
precisions of 2\% and $\sim$0.03 dex, respectively.
We also determine the radius, mass, and age with precisions of  
2--5\%, 7--11\%, and $\sim$35\%, respectively, using grid-based analyses.
Coupling the stellar parameters with photometric data 
yields an asteroseismic distance with a precision better 
than 10\%.
A \vsini\ measurement provides a 
rotational period-inclination correlation, and using the rotational 
periods from the recent literature, we constrain 
the stellar inclination for three of the stars.
An Li abundance analysis yields
an independent estimate of the age, but this is inconsistent with the asteroseismically 
determined age for one of the stars.
We assess the performance of five grid-based analysis methods 
and find them all to 
provide consistent values of the surface gravity to $\sim$0.03 dex
when both atmospheric and seismic constraints are at hand.
The different grid-based analyses all yield fitted values of radius and mass 
to within 2.4$\sigma$, and taking the mean of these results reduces
it to 1.5$\sigma$.
The absence of a metallicity constraint when the average large frequency separation
is measured with a precision of 1\% biases the fitted radius and 
mass for the stars
with non-solar metallicity (metal-rich KIC~11395018 and metal-poor
KIC~10273246), while 
including a metallicity constraint reduces the uncertainties
in both of these parameters by almost a factor of two.
We found that including the average small frequency separation improves
the determination of the age only for 
KIC~11395018 and KIC~11234888, and for the latter this improvement was  
due to the lack of strong atmospheric constraints.}
{}
{}
{}
{}

\authorrunning{Creevey et al.}
   \maketitle
%


\section{Introduction}

Stars like the Sun have deep convective envelopes where stochastic excitation gives rise
to a rich spectrum of resonant oscillation modes (e.g. \citealt{bg94,jcd05,aer10}). 
The frequencies of the modes depend on the journey that the waves make through the star, 
so that if the seismic signatures can be observed, they provide very accurate probes of the 
stellar interior \citep{ulr70,ls71,jcd07,met10,deh10a,dm10,vg10,cb11}. 
The oscillations have tiny amplitudes, 
and with photometry they can only 
be revealed with very long high-precision time series from space, e.g. with
CoRoT \citep{bag03,app08,gar09,mos09,deh10b,mat10a}.

The {\it Kepler} satellite \citep{bor10}, which was launched in early 2009, is providing photometric data of  
outstanding quality
on thousands of stars, see e.g. \citet{bed10}, \citet{gil10a}, \citet{hek10}, \citet{hub10}, \citet{kal10}, and \citet{ste10}.
With an average cadence of 30 minutes, the primary objectives of the mission, to search for and characterise Earth-like
planets,  can be reached.
For a smaller sample of stars ($\sim$512), the short cadence of one minute provides
the time-sampling adequate for detecting oscillation signatures present in solar-like stars, thus enabling a characterization of
such planet-hosting stars.

Due to the low amplitude of the oscillation modes, some of the individual frequencies ($\nu_{l,n}$ with
degree $l$ and radial order $n$)
may not 
be detectable.  
However, many analysis techniques allow one to determine seismic signatures in
relatively low signal-to-noise ratio (S/N) power spectra
\citep{hub09,ma09,rox09,cam10,hek10,kar10,mat10b}.
These signatures are primarily  {\sl i)} the average large frequency separation \mlsep\ where $\Delta\nu_{l,n} = \nu_{l,n} - \nu_{l,n-1}$, and {\sl ii)} the frequency corresponding to the maximum of the bell-shaped amplitude spectrum \numax\ (e.g. \citealt{hub10}).
If the individual frequencies are available, then \mlsep\ can be
determined with somewhat higher precision.
\citet{hek11} and \citet{ver11} compare \mlsep, \numax, and the uncertainties from a variety of 
established analysis techniques.

Anticipating the seismic quantities \mlsep\ and \numax\ to be measured 
for many stars,
various pipeline methods based on stellar evolution and structure models
have been developed to use these data to determine stellar 
properties, such as radius and mass, in an efficient and automatic manner.
In the ideal case, this type of grid-based approach provides good estimates of the 
parameters for a subsequent detailed seismic study. 
However, when oscillation frequencies are not available, grid-based
methods still provide reliable estimates of the mean density, surface gravity,
radius, and mass.
For example, Stello et al. (2009a) compared the radius determined from different
automatic analyses 
using simulated data and find that 
the radius can be determined with a precision of 3\%.
\citet{gai11} made a detailed study of grid-based methods for 
asteroseismology using simulated data and also solar data.  
They investigated the errors in the parameters and concluded that the 
surface gravity can be determined with practically no systematic bias.
Quirion et al. (2010) compared their automatic determination 
of stellar parameters with direct measurements of mass and radius for eight
bright nearby targets
(using interferometry and/or binaries), and
found agreement to within 1$\sigma$ for all stars except one.
Determining accurate stellar properties, in particular 
for single stars with $V \gtrsim 7$, is an important step towards
understanding stellar structure and evolution all across the HR diagram,
and seismic observations provide possibly the most 
accurate way of doing this.
This, in turn, can help in studies such as galactic stellar populations, e.g.
\citet{mig11,mos11}

Several stars observed by \kep\ were selected to be monitored at the
short-cadence rate for the full duration of the mission in order to test
and validate the time series photometry \citep{gil10b}.
In this paper we study five of these stars which 
show clear solar-like oscillation signatures. 
We use the global seismic quantities
 to determine their
surface gravity, radius, mass, and the age
while also assessing the validity of grid-based analyses.

The five  stars have the 
following identities from
the Kepler Input Catalog (KIC): KIC~11395018, KIC~10273246, 
KIC~10920273, KIC~10339342, and KIC~11234888, hereon referred to as 
C1, C2, C3, C4, and C5\footnote{Each of these stars has a pet cat name assigned to it within this 
collaboration.  These are Boogie, Mulder, Scully, Cleopatra, and Tigger, respectively.}, 
respectively, and their characteristics are given in Table~\ref{tab:stars}.
The {\it Kepler} short-cadence Q01234 (quarters 0 -- 4)
time series photometry provides the 
global seismic quantities.
The atmospheric parameters were determined from 
both
ground-based 
spectroscopic data --- analysed by five different
methods --- and photometric data (Sect.~2).
Five grid-based analysis methods based on stellar models 
are presented (Sect.~3) and used to 
determine \logg, the mean density, radius, mass, and age of each of the stars
(Sect.~4).
We also test the influence of an 
additional global seismic constraint, 
the mean small frequency separation \mssep\ where 
$\delta \nu_{0,n} =  \nu_{0,n}- \nu_{2,n-1}$.
Possible sources of systematic errors, such as the input
atmospheric parameters, and using different physics in the models are discussed
(Sect.~5) and then we 
assess the performance of each grid-based analysis method (Sect.~6).
In Sect.~\ref{sec:indep} we combine the stellar properties determined
in Sect.~4 
with published results
to provide an asteroseismic distance, to constrain the rotational period, and
to estimate the inclination of the rotation axis.
We also compare the asteroseismic age
to that 
implied from a lithium abundance analysis.

\begin{table*}
\caption{Basic data for the five solar-type stars.\label{tab:stars}}
\begin{center}
\begin{tabular}{llllllllll}
\hline\hline
{\it Kepler} ID& Adopted & Cat name & RA & DEC & $Kp$ & Time series&Spectral\\
&Name& &(hrs min sec) & ($^{\circ}$ ' ") &(mag) & (days) & Type$^{\star}$\\
\hline\hline
KIC~11395018 & C1 & Boogie &19:09:55 & 	49:15:04&10.762 & 252.71 & G4-5IV-V\\
KIC~10273246 & C2 & Mulder & 19:26:06 & 	47:21:30&10.903 & 321.68 &F9IV-V\\
KIC~10920273 & C3 & Scully &19:27:46 & 	48:19:45&11.926 & 321.68 &G1-2V\\
KIC~10339342 & C4 & Cleopatra &19:27:05 & 	47:24:08 &11.984 & 321.68 &F7-8IV-V\\
KIC~11234888 & C5 & Tigger &19:07:00 & 	48:56:07&11.926 & 252.71 & ... \\
\hline\hline
\end{tabular}
\end{center}
$^{\star}$ Spectral type determined from the ROTFIT method.
\end{table*}%

\section{Observations} 
\subsection{Seismic observations\label{sec:timeseries}}
The \kep\ targets C1, C2, C3, C4, and C5 have been observed at short cadence for at least eight months (Q0--4) since the beginning of \emph{Kepler} science operations on May 2, 2009. 
Observations were briefly interrupted by the planned rolls of the spacecraft and by three unplanned safe-mode events. The duty cycle over these approximately eight 
months of initial observations was above 90\%.
After 252 days, two CCD chips failed, which affected 
the signal for the targets C1 and C4.
The time series were analysed
using the raw data provided by the Kepler Science
Operations Center \citep{jen10},
subsequently corrected as described by \citet{gar11}. 
\citet{cam11} and \citet{mat11} presented details on the data calibration, as
well as an in-depth study of the time series
for stars C1, C2, C3, and C5 using a variety of documented analysis methods.
In this paper we use two methods to determine the global seismic quantities:
an automatic pipeline package A2Z \citep{mat10b} and a fit to the 
individual frequencies.

\subsubsection{Determination of seismic parameters using the A2Z package}

The A2Z pipeline looks for the periodicity of p modes in the power 
density spectrum (PDS) by computing the power spectrum of the 
power spectrum (PS2). 
We assume that the highest peak in the PS2 corresponds 
to $\langle \Delta \nu \rangle$/2. 
We then take a 600~$\mu$Hz-wide box in the PDS,  
compute its PS2 and normalise it by the standard deviation of the 
PS2, $\sigma$. 
We repeat this by shifting the box by 60~$\mu$Hz 
and for each box, we look for the highest peak in the range 
[$\langle \Delta \nu \rangle$/2 - 10~$\mu$Hz, 
$\langle \Delta \nu \rangle$/2 + 10~$\mu$Hz]. 
The boxes where the maximum power normalised by $\sigma$ is above the 95\% 
confidence level threshold delimit the region of the p modes, 
[$f_{\rm min}$,$f_{\rm max}$].   
The uncertainty on \mlsep\ is taken as the 
value of the bin around the highest peak in the PS2 computed by taking the 
PDS between $f_{\rm min}$ and $f_{\rm max}$. 

We estimate the frequency of maximum power, \numax, by 
fitting a Gaussian function  
to the smoothed 
  PDS between
$f_{\rm min}$ and $f_{\rm max}$.
The central frequency of the Gaussian is \numax, and
its uncertainty is defined as the smoothing factor, $1\times$\mlsep, 
resulting in a precision of between 6\% and 8\%.

\subsubsection{Determination of \mlsep\ from individual frequencies}

The frequencies of the oscillation modes for C1, C2, C3, and 
C5, have recently been published by 
\citet{mat11} and \citet{cam11}.
We list these frequencies in the appendix.
(For C4 the S/N in the power spectrum is too low to accurately determine
the frequencies.)
Following the approach by \cite{whi11}, 
we performed an unweighted linear least-squares fit
to the $l=0$ frequencies as a function of radial order
using the available range of frequencies [$f_{\rm min}$,$f_{\rm max}$]. 
The fitted gradient of the line is \mlsep, and 
the uncertainties are derived directly 
from the fit to the data. 

We used \mlsep\ determined from this method
together with \numax\ from the A2Z pipeline as the seismic data for 
C1, C2, C3, and C5.  For C4 we used both \mlsep\ and \numax\ from the 
A2Z pipeline.
Table~\ref{tab:seismicdata} lists the average seismic parameters, and these
are in agreement with those given in
\citet{cam11} and \citet{mat11}.

\subsubsection{Mean small frequency separation}
Because the individual frequencies are available, we
can readily calculate the mean small frequency separation \mssep,
which serves as an extra constraint on the stellar parameters.
However, calculating this quantity from the models implies 
a calculation of the theoretical oscillation frequencies for each
model, which is not the main purpose of 
most of the pipelines decribed here, and it is generally
an observable that is not available for stars with 
low S/N power spectra such as C4.
We derived \mssep\ from the individual frequencies within the 
[$f_{\rm min},f_{\rm max}$] range
and we list these values in Table~\ref{tab:seismicdata}, however, we only 
use it 
as a constraint on the models in Sect.~\ref{sec:mssep}.

\subsubsection{Solar seismic parameters}
We analysed the solar frequencies in the same way as the 
five \kep\ stars. 
We derive \mlsep\ = 135.21 $\pm$ 0.11 \mhz\ using the frequency
range [1000,3900] \mhz\ and the 
oscillation frequencies from \citet{bro09}, while
\citet{mat10b} derived \numax = 3074.7 \pmm\ 1.02 \mhz\ for the Sun.
The uncertainty in \numax\ is lower than 
1$\times$\mlsep, and
so we artificially increased the uncertainty to 145 \mhz\ (5\%) 
for a more
homogenous analysis, in line with its expected precision according 
to \citet{ver11}.
We also calculate \mssep = 8.56 \pmm\ 0.28 \mhz.

\begin{table}
\caption{Mean seismic parameters determined from the \kep\ data. \label{tab:seismicdata}}
\begin{center}
\begin{tabular}{llllllll}
\hline\hline
KIC ID&Star& \mlsep  & \numax &$f_{\rm min}$, $f_{\rm max}$&\mssep\\
&& (\mhz) & (\mhz) & (\mhz) & (\mhz)\\
\hline\hline
11395018&C1&47.52$\pm$0.15 & 834$\pm$50 & 686,972 &4.77\pmm0.23 \\
10273246&C2&48.89$\pm$0.09&838$\pm$50 & 737,1080 & 4.40\pmm0.44\\
10920273 &C3&57.27$\pm$0.13 & 990$\pm$60&826,1227 & 4.76\pmm0.14\\
10339342 &C4&22.50$\pm$1.50&324$\pm$25&219,437 & ...\\
11234888 &C5&41.81$\pm$0.09&673$\pm$50&627,837 & 2.59\pmm0.40\\
\hline\hline
\end{tabular}
\end{center}
\end{table}%

\subsection{Spectroscopic Observations}
We observed the targets C1 -- C4 using the FIES spectrograph
on the Nordic Optical Telescope (NOT) telescope located in 
the Observatorio del Roque de los Muchachos on La Palma.
The targets were observed during July and August 
2010 using the medium-resolution mode (R~=~46,000). 
Each target was observed twice to give total exposure times of 46, 46, 60, and 60 minutes respectively.
This resulted in an S/N of $\sim$80, 90, 60, and 60 in the wavelength region of 6069 -- 6076 \AA.
The calibration frames were taken using a Th-Ar lamp.
The spectra were reduced using FIESTOOL\footnote{http://www.not.iac.es/instruments/fies/fiestool/FIEStool.html}.

The reduced spectra were analysed by several groups
independently using the following methods:
SOU \citep{sou07,sou08}, VWA \citep{bru10}, ROTFIT \citep{fra06}, 
BIA \citep{sne73,bia11}, and NIEM \citep{np05}.
Here we summarise the main procedures for analysing the spectroscopic data,
and we refer readers to the appendix and the corresponding papers for a more detailed description of each method.

Two general approaches were taken to analyse the atmospheric spectra, 
both using the spectral region $\sim$4300--6680 \AA. 
The first involved measuring the equivalent widths (EW) of lines and then imposing 
excitation and ionization equilibrium using a spectroscopic analysis in local thermodynamic
equilibrium (SOU, BIA).
The second approach was based on directly comparing 
the observed spectrum with a library of synthetic spectra or reference stars (see,
e.g., \citealt{kat98,sou98}), either using full regions of the spectrum
(ROTFIT), or 
regions around specific lines (VWA, NIEM).
The atomic line data were taken from the Vienna Atomic Line Database 
\citep{kup99} and 
\citet{ch04}\footnote{http://www.user.oat.ts.astro.it/castelli/grids.html}.
The MARCS  \citep{gus08} and ATLAS9 \citep{kur93} model atmospheres were used, and the 
synthetic spectra were computed with the SYNTHE \citep{kur93}
and MOOG \citep{sne73} codes.
An 
automatic spectral type classification (cf. Table~\ref{tab:stars}) is given by ROTFIT.

\begin{figure}
\includegraphics[width = 0.49\textwidth]{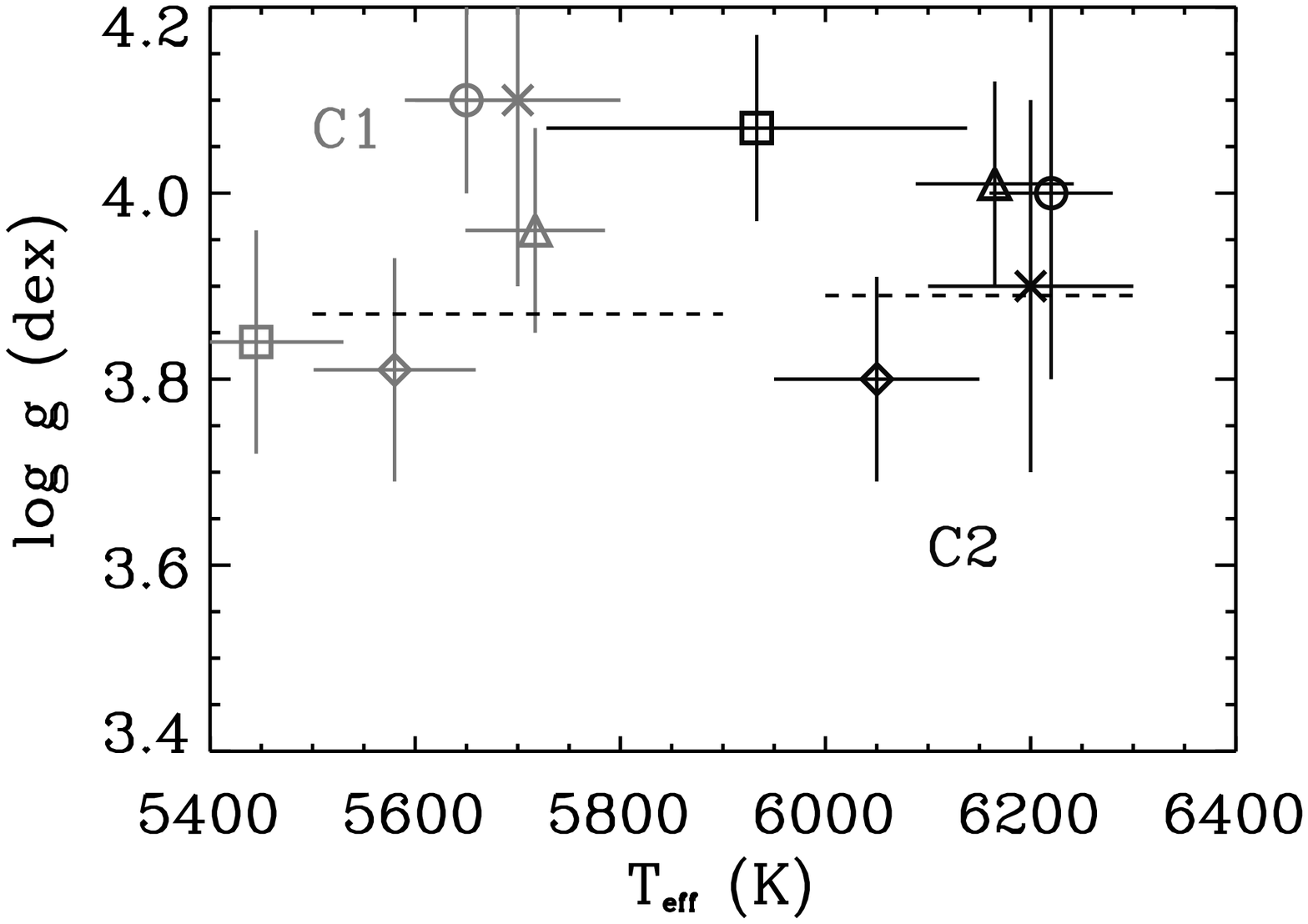}
\includegraphics[width = 0.49\textwidth]{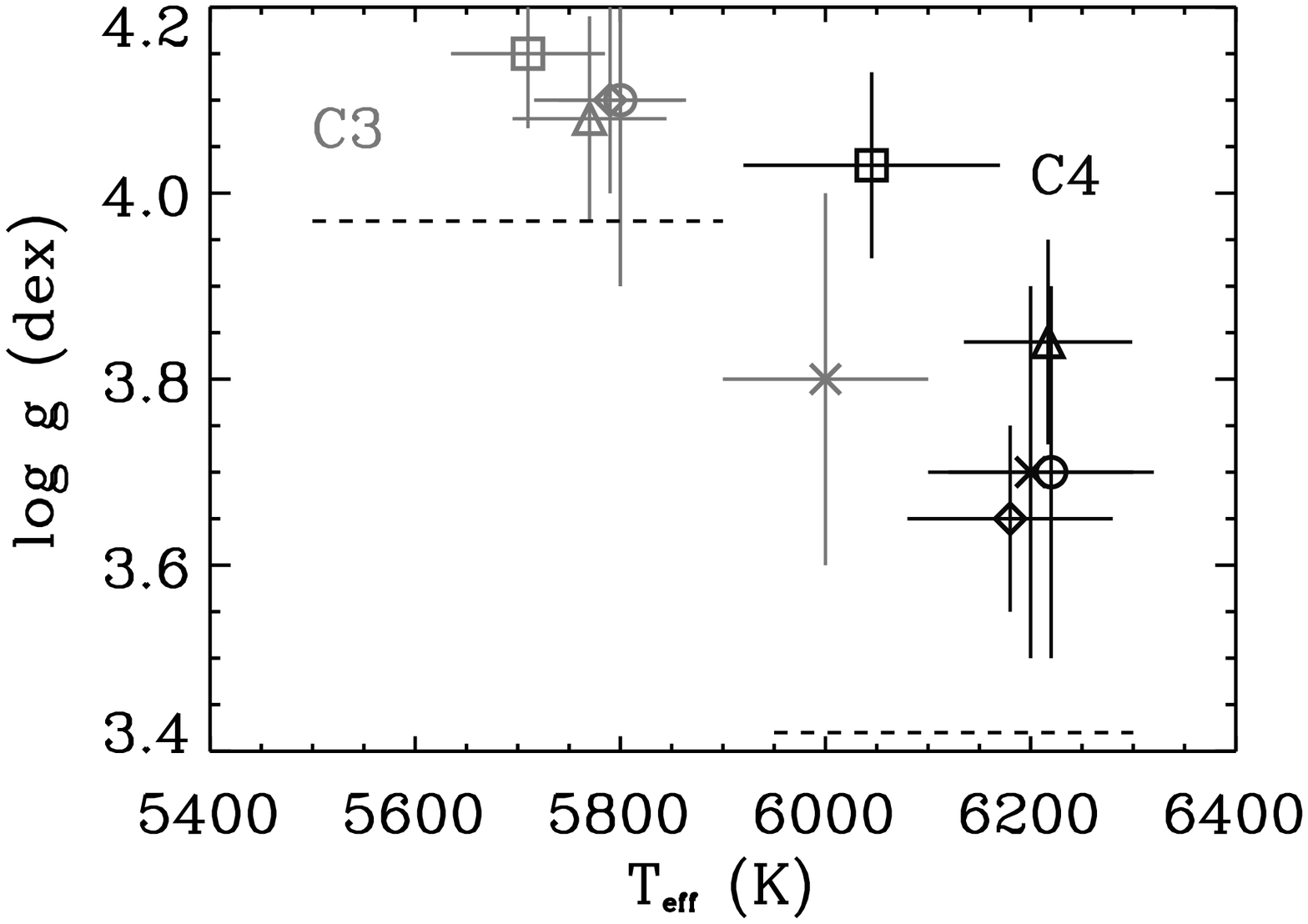}
\caption{\teff\ and \logg\ derived from five spectroscopic analyses.  
Each symbol represents the results from one analysis method: 
$\triangle$=SOU, $\square$=ROTFIT, $\diamond$=VWA,
$\circ$=BIA, $\times$=NIEM.  The dashed lines show the $\langle \log g_{\rm MR} \rangle$ 
values derived from the seismic data alone (see Sect.~\ref{sec:constlogg}).
\label{fig:tefflogg12}}
\end{figure}

 The derived atmospheric parameters for C1 -- C4 for each method are given
in Table~\ref{tab:atmos}, and 
Fig.~\ref{fig:tefflogg12} shows the fitted \teff\ and \logg\  for stars 
C1 and C2 (grey and black, respectively, top panel), 
and C3 and C4 (grey and black, respectively, bottom panel).
Each symbol represents the results from one spectroscopic analysis:
$\triangle$=SOU, $\square$=ROTFIT, 
$\displaystyle \diamond$=VWA,
$\circ$=BIA, $\times$=NIEM.
The figures emphasise the correlations between  
the two parameters, especially for C1 and C2:  
a lower \teff\ is usually fitted with a lower \logg.
The dotted lines represent the asteroseismic determination of \logg, as 
explained in Sect.~\ref{sec:constlogg}.
Considering the low S/N of the intermediate resolution spectra, we find 
that there is an overall good agreement (within 1--2$\sigma$) between the methods.
However, it is also clear from Table~\ref{tab:atmos} that there are some trends corresponding to the method used.
 For example,
the BIA method gives a systematically higher \teff\ than the VWA method,
the ROTFIT method generally yields higher \logg\ than the other methods, 
and the metallicity is on average higher when determined 
using EW methods (SOU, BIA) than with the line-fitting methods (NIEM, VWA, ROTFIT). 
A general comparison between the spectroscopic methods 
based on a much larger sample of stars with
high resolution spectra
is currently on-going, and will
be reported elsewhere.

\begin{table*}
\caption{Atmospheric parameters determined by various spectroscopic analyses of the same
NOT spectra, and from photometric analysis of the KIC photometry ('P/A').
\label{tab:atmos}}
\begin{center}
\begin{tabular}{lllllllllll}
\hline\hline
 & & \teff & \logg & \feh & $\xi_t$ & \vsini \\
 && (K) & (dex) & (dex) & (\kms) & (\kms)\\
\hline
C1\\
SOU && 5717$\pm$68&3.96$\pm$0.11&+0.35$\pm$0.05&1.30$\pm$0.03&...\\
ROTFIT & & 5445$\pm$85 & 3.84$\pm$0.12&+0.13$\pm$0.07&... & 1.1$\pm$0.8\\
VWA & & 5580$\pm$79 & 3.81$\pm$0.12 & +0.19$\pm$0.06 & 1.40$\pm$0.13&... \\
BIA &&5650$\pm$60&4.10$\pm$0.10&+0.36$\pm$0.10&1.20$\pm$0.20&... \\
NIEM& &5700$\pm$100 & 4.10$\pm$0.20& +0.13$\pm$0.10&0.70$\pm$0.40&... \\
P/A & & 5660$\pm$57&... &... &... &... \\
\\
C2\\
SOU && 6165$\pm$77&4.01$\pm$0.11&--0.04$\pm$0.06&1.48$\pm$0.05&... \\
ROTFIT && 5933$\pm$205&4.07$\pm$0.10&--0.21$\pm$0.08&...&3.2$\pm$1.5\\
VWA &  & 6050$\pm$100 & 3.80$\pm$0.11 & --0.18$\pm$0.04 & 1.50$\pm$0.10&...  \\
BIA && 6200$\pm$60&4.00$\pm$0.20&--0.04$\pm$0.07&1.50$\pm$0.20&... \\
NIEM && 6200$\pm$100 & 3.90$\pm$0.20& --0.18$\pm$0.05&0.50$\pm$0.40&... \\
P/A && 6380$\pm$76&... &... &... &... \\
\\
C3\\
SOU && 5770$\pm$75&4.08$\pm$0.11&+0.04$\pm$0.05&2.11$\pm$0.08&... \\
ROTFIT&&5710$\pm$75&4.15$\pm$0.08&--0.02$\pm$0.07&...&1.5$\pm$2.2\\
VWA & & 5790$\pm$74 & 4.10$\pm$0.10 & --0.04$\pm$0.10 & 1.15$\pm$0.10&...  \\
BIA && 5800$\pm$60&4.10$\pm$0.20&+0.03$\pm$0.07&1.20$\pm$0.20&... \\
NIEM && 6000$\pm$100 & 3.80$\pm$0.20& --0.03$\pm$0.08&1.00$\pm$0.40&... \\
P/A && 5880$\pm$53&... &... &... &... \\
\\
C4\\
SOU&&6217$\pm$82&3.84$\pm$0.11&--0.11$\pm$0.04&1.60$\pm$0.20&... \\
ROTFIT && 6045$\pm$125&4.03$\pm$0.10&--0.23$\pm$0.08&...&4.0$\pm$2.8\\
VWA & & 6180$\pm$100 & 3.65$\pm$0.10 & --0.15$\pm$0.10 & 1.75$\pm$0.10&...  \\
BIA &&6200$\pm$100&3.70$\pm$0.20&--0.06$\pm$0.08&1.60$\pm$0.20&... \\
NIEM && 6200$\pm$100 & 3.70$\pm$0.20& --0.17$\pm$0.06&0.50$\pm$0.40&... \\
P/A&&6280$\pm$63&... &... &... &... \\
\\
C5\\
P/A &&6240$\pm$60&... &... &... &... \\
\hline\hline
\end{tabular}
\end{center}
\end{table*}

\subsection{Photometrically derived atmospheric parameters\label{sec:marc}}
Ground-based SLOAN-$griz$ photometry 
is available for a large number of stars in the field of view.  
The Kepler Input Catalog (KIC) lists the magnitude in the wide Kepler band pass, $Kp$, 
as well as the $griz$ magnitudes and the 
stellar parameters (\teff, \feh, \logg, radius \rad) derived using these data.
However, the primary purpose of the KIC was to allow 
discrimination of dwarfs from other classes of stars to aid in the selection of 
 planet-hosting candidates.
It has become clear since the time series data became available 
that 
the KIC \teff\ are not always accurate on a star-to-star basis
\citep{mol10a,mol10b,leh11}. 
\teff\ were, therefore, re-calculated by Pinsonneault \& An (2011) using 
SLOAN photometry and the YREC models \citep{an09,dem08}, and cross-checking the 
results using the infra-red flux method 
calibration based on 2MASS photometry \citep{cas10}.
The \teff\ and uncertainties are listed in Table~\ref{tab:atmos} with the heading 'P/A'.

\section{Seismic methods\label{sec:pipeline}}

The nearly uninterrupted short-cadence \kep\ time series 
yield power spectra that exhibit some signatures of oscillations.
Even from low S/N power spectra, the global 
seismic quantities \mlsep\ and  \numax\  
can easily be determined without having to measure individual frequencies.
In preparation for the hundreds of stars in the \kep\ data where 
these seismic parameters are readily available, 
several  pipeline codes have been developed to use stellar models to 
 infer the surface gravity \logg, mean density \md, 
radius \rad, mass \mass, and age \age\ of stars from the 
 global seismic quantities 
supplemented with atmospheric parameters, such as \teff. 
In this analysis we used the following methods which are discussed below: 
CESAM2k/Mumbai \citep{maz05},
Yale-Birmingham \citep{bas10, gai11},
RADIUS \citep{ste09a,met10}, 
SEEK \citep{qui10}, 
and
RadEx10 (Creevey in prep.).
We used five different analysis methods in
order to test the validity of our results, and 
to assess the performance of each code for producing
reliable stellar parameters.

\subsection{CESAM2k/Mumbai}

The analysis using the CESAM2k stellar evolution code \citep{ml08} is based 
on the comparison of both seismic and non-seismic observations ($q$=\{\mlsep,\numax,
\teff,\logg,\feh,\mssep\})
with those calculated from a grid of stellar evolution models. 
This version of CESAM2k uses the OPAL equation of state \citep{rn02} and 
the OPAL opacities \citep{ir96}
supplemented by the low temperature opacities of \citet{af94}.
The solar mixture is given by \cite{gs98} and the
NACRE nuclear reaction rates \citep{ang99} are used.
Convection is
described by the standard mixing length theory \citep{bv58}. The
models also include microscopic diffusion of helium and heavy elements,
following the prescription of \citet{pm91} for masses $\le 1.3$ \msol. 

The grid of models used in this analysis spans the mass range of
0.80~\msol\ to 1.70~\msol, in steps of 0.02 \msol. 
The
initial metallicities of the models $Z_{\rm i}$ range from 0.005 to 0.030 in steps of 0.005, 
and for each value of $Z_{\rm i}$, five
different combinations of ($X_{\rm i}$, $Y_{\rm i}$) are used, where $Y_{\rm i}$ is the initial helium fraction. 
Three
mixing length parameters $\alpha=1.8,1.9,2.0$ are considered, and for
stars with convective cores, we used convective overshoot to the extent of $\alpha_{\rm ov} \times H_{\rm P}$ ($H_{\rm P}$ = pressure scale height), with
$\alpha_{\rm ov}$ = \{$0.00, 0.05, 0.10, 0.15, 0.20, 0.25$\}. 
The stellar evolutionary tracks start from
the zero age main sequence (ZAMS) and continue until $\log (T_{\rm eff}) \approx
3.715$, which corresponds to on or near the red giant branch.
For each model during the evolution, 
oscillation frequencies for low degree modes ($\ell=0,1,2,3$) are
computed under the adiabatic approximation using the Aarhus pulsation
package, ADIPLS \citep{jcd08b}. 

For every star, the average large frequency
separation was computed between the
same frequency limits as used in the observed data for the corresponding
star and run. The average was calculated by integrating the individual
frequency separation as a 
function of frequency between the given limits and dividing by the
frequency range. The frequency of maximum amplitude, \numax, has been 
calculated using the scaling relation provided by \citet{kb95}:
\begin{equation}
\nu_{\rm max} \simeq \frac{M/M_{\odot}}{(R/R_{\odot})^2\sqrt{T_{\rm eff}/5777\,{\rm K}} }3,050 \mu{\rm Hz}
\label{eqn:numax}
\end{equation}

We constructed a reduced $\chi^2_R$ using the available data as follows:
\begin{equation}
\chi^2 = \sum_{i=1}^{n_\mathrm{par}}
\left(\frac{q_{i,\mathrm{obs}}-q_{i,\mathrm{mod}}}{\sigma_{i,\mathrm{obs}}}\right)^2,
\label{eqn:chi2}
\end{equation}
where the sum is over all the available constraints $q_i$, 
and $\chi^2_R$ takes into account the number of constraints. 
The subscripts
``obs'' and ``mod'' refer to the observed and model values,
respectively. The quoted uncertainty in the $i$th observed data is denoted by
$\sigma_{i,\mathrm{obs}}$.
We sought the minima of $\chi^2_R$ to provide estimates of the
stellar parameters. The quoted values of the parameters and their uncertainties 
are the midpoints and half the span of the
ranges 
of parameters for which $\chi^2_R \leq 1$.

\subsection{Yale-Birmingham}
The Yale-Birmingham \citep{bas10} code as described by
\citet{gai11} is a grid-based method for determining a star's mass, radius, and age\footnote{In the absence of a metallicity measurement the determination of
the age is hindered, see \citet{gai11}}.
The grid of models was constructed using  the Yale Rotation and Evolution Code 
\citep{dem08} in its non-rotating configuration. 
The input physics includes the OPAL equation of state tables,
the OPAL temperature opacities 
supplemented with low temperature opacities from \citet{fer05}, and the NACRE 
nuclear reaction rates. 
 All models include gravitational 
settling of helium and heavy elements using the formulation of \citet{tho94}. We use
the Eddington $T(\tau)$ relation (here $\tau$ means optical depth), and the adopted mixing length parameter is $\alpha = 1.826$. An 
overshoot of $\alpha_{\rm ov} = 0.2$ was assumed for models with convective cores. 
The grid consists of 820,000 individual models
with masses ranging from 0.8 to 3.0 $M_{\odot}$ in steps of 0.2$M_{\odot}$.
These models have
[Fe/H] ranging from +0.6 to $-0.6$ dex in steps of 0.05 dex. We
assume that [Fe/H] = 0 corresponds to the solar abundance ($Z_{\odot}/X_{\odot}=0.023$)
as determined by \citet{gs98}.
The model value for \numax\ is calculated using 
Eq.~(\ref{eqn:numax}), and 
\mlsep\  is determined using
\begin{equation}
\frac{\langle \Delta \nu \rangle}{\langle \Delta \nu \rangle_{\odot}} \simeq
\sqrt{\frac{\rho}{\rho_{\odot}}} = 
\left( \frac{M}{M_{\odot}}\right)^{\frac{1}{2}} \left( \frac{R}{R_{\odot}} \right)^{-\frac{3}{2}}
\label{eqn:dnu}
\end{equation} where  $\langle \Delta \nu \rangle_{\odot}$ = 134.9 \mhz\ (see \citealt{kb95}).

The Yale-Birmingham pipeline finds the maximum
likelihood of a set of input parameters calculated with respect to
the grid of models.  The estimate of the parameter is obtained by taking
an average of the points
that have the highest likelihood. We average all points with
likelihoods over 95\% of the maximum value of the likelihood 
functions. 
For a given observational (central) input parameter
set, the first key step in the method is generating 10,000 input
parameter sets by adding different random realizations of Gaussian
noise to the actual (central) observational input parameter set. The
distribution of any parameter, say radius, is obtained from the central parameter set and the
10,000 perturbed parameter sets form the distribution function. The final
estimate of the parameter is the median of the distribution.
 We used 1$\sigma$ limits from the median as a measure of the uncertainties.

The likelihood function is formally defined as
 \begin{equation}
 \mathcal {L}=\left(\prod^{n}_{i=1}\frac{1}{\sqrt{2\pi}\sigma_{i}}\right)\times
 \exp(-\chi^{2}/2), \label{eq:likelihood}
 \end{equation}
where $\chi^2$ is given in Eq.~\ref{eqn:chi2}, and
$q$ $\equiv$ \{$T_{\rm eff}$, [Fe/H], $\Delta\nu$, $\nu_{\rm max}$\}.
From the form of
the likelihood function in Eq.~\ref{eq:likelihood} it is
apparent that we can easily include more inputs, or drop some
inputs depending on the availability of data.

\subsection{RADIUS}

The RADIUS pipeline \citep{ste09a} is based on a large grid of ASTEC models
\citep{jcd08a} using the EFF equation of state
\citep{egg73}. We used the opacity tables of \citet{ri95}
and \citet{kur91} (for $T<10^4$\,K), with solar mixture of
\citet{gn93}. Rotation, overshooting, and diffusion were not
included. The grid was created with fixed values of the mixing-length
parameter, $\alpha=1.8$, and the initial hydrogen abundance of
$X_{\mathrm{i}}=0.7$.  The resolution in $\log (Z/X)$ was 0.1 dex 
between
$0.001<Z<0.055$, and the resolution in mass was $0.01\,M_\odot$ from 0.5 to
$4.0\,M_\odot$.  The evolution begun at the ZAMS and continued to the tip
of the red giant branch.  To convert between the model values of $Z$ and
the observed [Fe/H], the pipeline used $Z_\odot = 0.0188$ \citep{cox00}.

Each output parameter was determined by selecting the set of models
that were within $\pm 3 \sigma$ of the observed input data.
We pinpointed a single best-fitting model using a $\chi^2$ formalism
and the 1$\sigma$ uncertainty is estimated as 1/6 of the maximum range of 
the values of the selected models.
The pipeline as described in detail by \citet{ste09a} had 
some slight modifications; for example,
the large frequency separation was
derived by scaling the solar value (see Eq.~\ref{eqn:dnu}) instead of
calculating it directly from the model frequencies.

\subsection{SEEK}
The SEEK procedure \citep{qui10} also makes use of a large grid of stellar models computed with the ASTEC
code. 
This version of ASTEC uses 
the OPAL equation of state \citep{rog96} along with the OPAL plus Ferguson \& 
Alexander opacity tables \citep{ir96,af94}, the element 
to element ratios in the metallic mixture of \citet{gs98}, 
and convection is 
treated with the mixing-length formulation of \citet{bv58}; the mixing length to 
pressure scale height ratio $\alpha$, characterizing the convective efficacy, is treated as a variable 
parameter in the SEEK fits. Neither diffusion nor overshooting is included. 
Oscillation frequencies for each model are calculated using the ADIPLS \citep{jcd08b} code.

Two subgrids were created:
the first subgrid comprises 
tracks with all combinations of 
$Z$ = [0.005, 0.01, 0.015, 0.02, 0.025, 0.03], $X_{\rm i}$ = [0.68, 0.70, 0.72, 0.74], and 
$\alpha$ = [0.8, 1.8, 2.8] while the second subset has $Z$ = [0.0075, 0.0125, 0.0175, 0.0225, 0.0275],
$X_{\rm i}$ = [0.69, 0.71, 0.73], $\alpha$ = [1.3, 2.3]. Every subset is composed of 73 tracks spanning from 
0.6 to 1.8 \msol\ in steps of 0.02 \msol\  and 
from 1.8 to 3.0 \msol\ in steps of 0.1, and each track was evolved until just after the base of the 
giant branch or $\tau$ = 15$\times 10^9$ yrs.
The metallicity combinations correspond to --0.61$\leq$ [Fe/H] $\leq$ 0.20.

SEEK compares an observed star with every model of
the grid and makes a probabilistic assessment of the stellar parameters, 
with the help of
Bayesian statistics.
Its aim is to draw the contour of good
solutions which is located around $\chi^2_R$. 
The priors used in
that assessment are flat for the age, the metallicity, the initial
helium ratio, and the mixing length parameter. The only non-flat prior is
related to the initial mass function and makes use of the \citet{cha01}
 IMF model, where $\xi(M) = 0.019 M^n$, with $n = -1.55$ for 
$M \leq 1.0 M_{\odot}$ and $n = -2.70$ when $M >1.0M_{\odot}$.
The details of the SEEK procedure, including the choice of priors, and an introduction to 
Bayesian statistics can be found in \citet{qui10}.

\subsection{RadEx10}

RadEx10 is a grid-based approach to determining the radius, mass, and age using some 
or all
of the following as input \{\mlsep,\numax,\teff,\logg,\feh\}.
It is based on the ASTEC code and uses 
the EFF equation of state of \citet{egg73} without Coulomb corrections,
the OPAL opacities \citep{ir96} supplemented by Kurucz opacities
at low temperatures, and solar
mixture from \citet{gn93}.
  The nuclear reaction rates came from \citet{bp92}, convection
is described by the mixing-length theory of \citet{bv58}, convective core
overshooting is included with $\alpha_{\rm ov}$ set to 0.25, and diffusion effects are ignored.

The grid considers models with masses from 0.75 -- 2.0 \msol\ in steps of 0.05 \msol, ages from ZAMS to
subgiant, $Z_{\rm i}$ spans 0.007 -- 0.027 in steps of $\sim0.003$, while 
 $X_{\rm i}$ is set to 0.70: this corresponds to  $Y_{\rm i} = 0.263 - 0.283$. 
 The mixing length parameter $\alpha = 2.0$ is used, which
 was  obtained by calibrating  the solar data.
To obtain the stellar properties of mass, radius, and age, we perturb the 
observations using a random Gaussian distribution, and compare the perturbed 
observations to the model observables to select an optimal model.
The \rad, \mass, \age, and their uncertainties are defined as the mean value of the 
fitted parameter from 10,000 realizations, with the standard deviations 
defining the 1$\sigma$ uncertainties.

\section{Stellar properties\label{sec:sectstellarproperties}}

\subsection{Constraining \logg\ with seismic data\label{sec:constlogg}}
The quantity \mlsep\ is proportional to the mean density of the star 
(see eq.~[\ref{eqn:dnu}]) 
and \numax\ also scales with \rad\ and \mass\ (see eq.~[\ref{eqn:numax}]). 
By making an assumption that these stars have roughly solar \teff\ 
with a large uncertainty, 
the seismic data alone should give a robust estimate of \logg\ by using the
scaling relations for \numax\ and \mlsep\ and solving for \rad\ and \mass.
The first two data columns in Table~\ref{tab:logg} show 
\logg\ and its uncertainty, denoted by the subscript '$\nu$',
when we use the seismic data and 
a \teff\ estimate of 6000 K \pmm\ 500.

The grid-based methods described in Sect.~\ref{sec:pipeline} also used
\mlsep\ and \numax\ from Table~\ref{tab:seismicdata}
as the only input observational data to their codes to obtain 
a {\it model asteroseismic} value of \logg. 
In this case 
we restricted the model \teff\ to less than 8000 K.
Table~\ref{tab:logg} shows the mean 
value of \logg, $\langle \log g_{\rm MR} \rangle$,
obtained by combining the results from the five methods.
We indicate that these are model-determined values by the subscript 'MR'.
As can be seen, there is very good agreement between the 
model-independent and model-dependent values of \logg.

Figure~\ref{fig:comparelogg} shows the difference between 
the asteroseismic \logg\ value returned by each 
method, $\log g_i$, and $\langle \log g_{\rm MR} \rangle$.
From left to right on the x-axis we show the results from
SEEK, RADIUS, RadEX10, Yale-Birmingham, and CESAM2k/Mumbai
(note the abbreviated labelling on the x-axis).
For each seismic method, we show the differences in \logg\ from
left  to right for stars C1, C2, C3, C4, C5, and the Sun.
For the CESAM2k/Mumbai method `CMum`, we show results for 
C1, C2, and C3 only, and these are labelled accordingly to avoid
confusion.

The fitted \logg\ values for each star and each method differ by
$\sim$0.05 dex, excluding C4, with 
the largest differences seen primarily 
between the RADIUS and Yale-Birmingham methods, 
but still matching within 2$\sigma$.
For C4 SEEK and RadEx10 give the largest dispersions, this difference
reaching 0.13 dex.  However, C4 is a very evolved MS star (\logg$\sim$3.50 dex)
and the results for RadEx10 are biased, since this grid concentrates
mostly in the MS and just beyond. 
If we ignore the result for RadEx10, we find a maximum difference between 
the results of 0.07 dex.
The differences of 0.05 and 0.07 dex are still in much better
agreement than the spectroscopic methods.

We also see that the statistical 
errors given by each method vary between 0.01 -- 0.05
dex.  However, all of the results fall to within 2$\sigma$ of the 
mean value.
Each reported uncertainty is not underestimated, but some grids take into
account some other variables, e.g. different mixing-length parameters, that 
will increase the reported value. 
For example, the uncertainties reported by Yale-Birmingham and
RADIUS are indeed correct, however,
because their grids are based on different physics and sets of parameters,
they obtain results that differ by more than 1$\sigma$.
Gai et al. (2010) did a systematic study of the uncertainties 
in stellar parameters determined by grid-based methods, including
the Yale-Birmingham grid. 
Their study included a detailed investigation of the errors in $\log g$,
and they found typical statistical uncertainties of $~0.014$ dex when \teff\
is included, in
agreement with those reported here. 
They also tested the systematic errors by using different 
grids and different parameters (e.g. a different mixing-length parameter)
and found that for stars similar to those used in this paper
i.e. with similar  \mlsep,
the half-width at half-maximum of the distribution of 
systematic errors is around
 $3$\% in \logg\ (see their Fig.~19, panel b). For a $\log g$ value of around
3.8, this would imply $\sigma\leq 0.1$ dex, in agreement with the
differences in \logg\ reported above.

As another example, the CESAM2k/Mumbai method reports results 
that take into account different values of the convective core overshoot
parameter in the models.  
The inclusion of more parameters/physics will yield
a more conservative error.  

While distinguishing between the statistical uncertainties and the 
systematic errors is not always clear, we can be confident
that combining the results from different grids provides 
a reliable determination of \logg\ while the dispersion among
the results is representative of a typical systematic
error found by using different sets of physics and parameters.
In order to provide accurate results with a 
conservative error we adopt the results from SEEK, whose
uncertainty is larger than the dispersion among the fitted
results.
In Table~\ref{tab:logg}, last three columns, we give
an estimate of the systematic error,
$\sigma_{\rm sys} = {\rm max}\{|\log g_{\rm SEEK} - \log g_i|\}$, 
and the \logg\ values and uncertainties provided by SEEK.

\begin{table}
\begin{center}
\caption{Surface gravity obtained from \mlsep\ and \numax\
using scaling relations and stellar models, denoted by subscripts '$\nu$' 
and 'MR', respectively.
\label{tab:logg}}
\begin{tabular}{lccccccccccccrrlllllll}
\hline\hline
Star &\logg$_{\nu}$&$\sigma_{\nu}$&$\langle \log g_{\rm MR} \rangle^{a}$&$\sigma_{\rm sys}$& \logg$_{\rm MR}^{b}$ &$\sigma_{\rm stat}^{b}$  \\
& (dex)& (dex)& (dex)& (dex)& (dex)\\
\hline
C1 &3.88&0.02&3.88 & 0.05    &3.87 &0.07      \\
C2 &3.88&0.03&3.89 & 0.04 &3.89 &0.06      \\
C3 &3.95&0.03&3.98 & 0.05   &3.97 &0.06     \\
C4 &3.47&0.04&3.49$^{c}$ & 0.13$^{c}$ &3.42 &0.06 &\\
C5 &3.78&0.04& 3.82 & 0.05   &3.80 & 0.07 &    \\
Sun&...&...& 4.42 & 0.02 & 4.43 & 0.02      \\
\hline\hline
\end{tabular}
\end{center}
$^a \langle \log g_{\rm MR} \rangle$ is the mean value of \logg\ provided
by all of the model results.\\
$^b$\logg$_{\rm MR}$ and $\sigma_{\rm stat}$ are the SEEK values.\\
$^{c}$ Discarding the value from RadEx10 yields 
$\langle \log g \rangle$ = 3.46 and $\sigma_{\rm sys}$ = 0.07 dex.
\end{table}

\subsection{Radius and mass}

The atmospheric parameters \teff\ and \feh\ are needed to derive the radius
and mass of the star.
Using the \logg\ values obtained in Sect.~\ref{sec:constlogg} we selected
one set of spectroscopic constraints as the optimal atmospheric parameters,
and to minimise the effect of the correlation of spectroscopically derived
parameters.
We chose the set whose \logg\ values matched closest
globally to the asteroseismically determined ones.
By inspecting Tables~\ref{tab:atmos} and \ref{tab:logg} we found that
the VWA method has the overall closest results in \logg.
We therefore combined these spectroscopic 
data with \mlsep\ and \numax\ from Table~\ref{tab:seismicdata}
and used these as the observational input data for the seismic analysis.  
For C5 we used the photometric
\teff.

\begin{figure}
\includegraphics[width=0.50\textwidth]{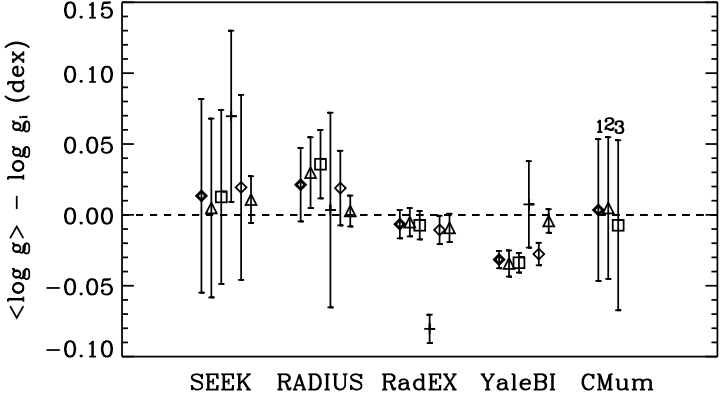}
\caption{Comparison of each seismic method's
asteroseismic values of \logg\ 
 to the
mean value $\langle \log g_{\rm MR} \rangle$ of the four or five methods.  
Each method is abbreviated and labelled on the x-axis, and for each 
method the results are shown from left to right for C1, C2, C3, C4, C5, and the 
Sun. The CMum (CESAM2k/Mumbai) method shows results for C1, C2, and C3 only.
\label{fig:comparelogg}}
\end{figure}

\begin{figure}
\includegraphics[width=0.50\textwidth]{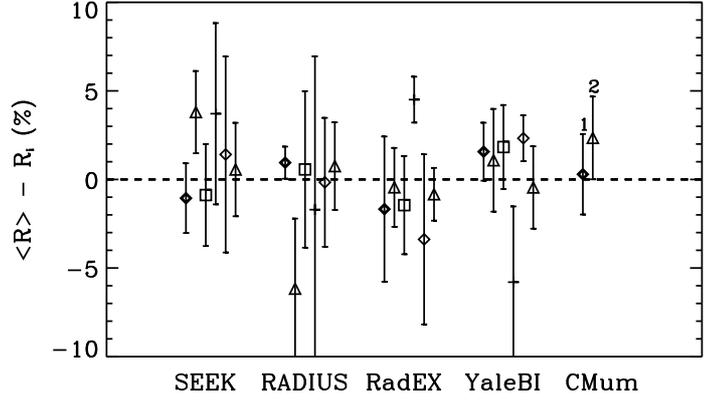}
\caption{Comparison of each seismic method's determination of \rad\ with
the mean value $\langle R\, \rangle$  using
both spectroscopic and seismic constraints.
See caption of Fig.~\ref{fig:comparelogg} for details.\label{fig:arad}}
\end{figure}

In Fig.~\ref{fig:arad} we show the deviation of the fitted
radius of each method $R_i$ from the mean value $\langle R\, \rangle$
in units of \%, 
with the mean radius in units of \rsol\ given in Table~\ref{tab:radius}.
The representation of the results is the same as in Fig.~\ref{fig:comparelogg} i.e.
from left to right for each method, we show C1, C2, C3, C4, C5, and the Sun.
Most of the results are in agreement with the mean value at a level of
1$\sigma$, and the uncertainties vary between approximately 1 and 4 \%
(see Sect.~\ref{sec:constlogg}).
To be consistent, we adopt the SEEK method as the reference one.
This choice is also justified by the following reasons:
i) SEEK covers the largest parameter space in terms of mass, age, metallicity, 
and mixing-length parameter,
ii) it has also been tested with direct measurements of mass and radius of
nearby stars, 
iii) and it determines a best
model parameter for each property 
(e.g. luminosity, initial metal mass fraction, and uncertainties),
which allows us to make further inferences as well as investigate the 
systematics.
In addition, it does not use the scaling relations, which may
introduce a systematic bias on the order of 1\% 
in the stellar parameters e.g. \citet{ste09b}.
We used the results from the other pipelines as a test of the systematic errors.

\begin{table}
\begin{center}
\caption{Radius determination from the SEEK method (first half of table) 
and combining all 
methods (second half of table) using spectroscopic and
asteroseismic data. 
\label{tab:radius}}
\begin{tabular}{cccccccccclllllll}
\hline\hline
Star & \rad&$\sigma$&$\sigma$ & $\langle R\, \rangle$&$\sigma_{\rm sys}$ &$\sigma_{\rm sys}$ 
& $\sigma_{\rm sys}/\sigma_i$  \\
& (\rsol) &(\rsol) & (\%) & (\rsol) &(\rsol)& (\%) &\\
\hline
C1  &2.23 &0.04 &2 &2.21 & 0.06  &3  & 2.3  \\
C2  &2.11 &0.05 &2 &2.19$^{a}$ & 0.22$^{a}$  &11 &2.4   \\
C3  &1.90 &0.05 &3 &1.88 & 0.07  &4   &1.2 \\
C4  &3.81 &0.20 &5&3.95$^{b}$ & 0.38 & 10&2.2\\
C5  &2.44 & 0.14 &6& 2.47 & 0.12 &5  &  1.0 \\
Sun& 0.99 & 0.03 &3& 1.00 & 0.01 &1    &1.0 \\
\hline\hline
\end{tabular}
\end{center}
$^{a}$ discarding the RADIUS result yields a value of 2.15 \rsol\ \pmm\
0.09.\\
$^{b}$ discarding the value from RadEx10 
yields $\langle R\, \rangle = 4.0$.
\end{table}

In Table~\ref{tab:radius} we list the \rad\ values found using SEEK, the 
uncertainties $\sigma$, the mean value of \rad\ obtained by
combining the results from all of the grids $\langle R\, \rangle$, 
$\sigma_{\rm sys} = {\rm max}\{|R_{\rm SEEK} - R_i|\}$ given in 
units of \rsol, and $\sigma_{\rm sys}$ normalised by $\sigma_i$ i.e.
${\rm max}\{|(R_{\rm SEEK} - R_i)/\sigma_i|\}$.
Using SEEK with its reference set of physics, we determine the radius of each
star with a typical statistical precision of 3\%.  
If we use the results from the other pipelines as a measure of the 
systematic error, then we report an accuracy in the radius of between 3 and 
5\% for C1, C3, C5, and the Sun.
The RADIUS method reports a radius that differs by 11\% from the 
radius of the SEEK method for C2, but this difference
corresponds to only 2.4$\sigma$ from the fitted \rad.
Without this value $\sigma_{\rm sys}$ reduces to 5\%.
However, we have no justification to remove this value or discard the 
possibility that this is closer to the correct value.
We remove the RadEx10 value
for C4, but this results in an insignificant change ($<1$\%). 
The method responsible for the largest deviation from the SEEK results for C4 
is the
Yale-Birmingham method, although they agree just above their 1$\sigma$ level.
In the final column of the table we show the maximum deviation of the
fitted radius from the SEEK radius but in terms of the uncertainty given
by each pipeline method $\sigma_i$.  
Here we see that all of the results are consistent with SEEK to 2.4$\sigma$.

Table~\ref{tab:amass} reports the values of the mass 
determined by SEEK, and the other seismic methods, 
using the same format as Table~\ref{tab:radius}.
The statistical uncertainties are of the order of 8\%, and 
the systematic errors are typically 5-12\%, with 
each grid-based method reporting the same fitted mass to within
2.4$\sigma$.
The fitted mass is highly correlated with the fitted radius, and thus
most of the same trends are found among each pipeline method 
for both radius and mass.
A much larger uncertainty in mass is reported for C5, and this is 
due to the lack of a metallicity constraint.

\begin{table}
\begin{center}
\caption{Mass determination from the SEEK method (first half of table) 
and combining all 
methods (second half of table) using spectroscopic and
asteroseismic data. 
\label{tab:amass}}
\begin{tabular}{cccccccccclllllll}
\hline\hline
Star& \mass &$\sigma$&$\sigma$ & $\langle M\, \rangle$ &$\sigma_{\rm sys}$ &$\sigma_{\rm sys}$ 
& $\sigma_{\rm sys}/\sigma_i$  \\
& (\msol) &(\msol) & (\%) & (\msol) &(\msol)& (\%) & \\
\hline
C1  &
1.37 &     0.11  &     8&  1.35 &  0.09 & 7 & 2.0\\
C2  &
 1.26&  0.10 &   8 &  1.37$^{a}$ &  0.43$^{a}$&   34 & 2.4\\
C3  &
1.25&     0.13&  11&  1.23& 0.06&   5 & 1.0\\
C4  &
1.79&   0.12&     7&  1.84$^{b}$&   0.09$^{b}$ &  5 & 1.5$^{b}$\\
C5  &
 1.44& 0.26& 18& 1.50&   0.18 &   12 & 1.6\\
Sun& 
0.97&0.06&  7&  1.01&0.08&9&1.0\\
\hline\hline
\end{tabular}
\end{center}
$^{a}$ discarding the RADIUS result yields a value of 1.32 \rsol\ \pmm\
0.14.\\
$^{b}$ discarding the value from RadEx10. 
\end{table}

\subsection{\logg\ and $\langle \rho \rangle$ using combined seismic and spectroscopic data \label{sec:loggrhomr}}
Combining \mlsep\ and \numax\ with the VWA spectroscopic values of \teff\ and 
the photometric value for C5, 
we calculated model-independent values of \logg\ just as explained in 
Sect.~\ref{sec:constlogg}. 
We also calculated the mean density of the star using \mlsep.
Table~\ref{tab:loggmeand} lists these properties, again denoted by '$\nu$' to
indicate that they are obtained directly from the data.
We also give the model-dependent values of 
\logg$_{\rm MR}$ and $\langle \Delta \nu\, \rangle_{\rm MR}$
as reported by SEEK.

Both the \logg\ and \md\ values are in good agreement when derived
directly from the data and when using stellar models. 
For \logg\ there 
is agreement to 1.5$\sigma$, and for \md\ the agreement is within 2.5$\sigma$.
These values represent a relative precision of 2\% for \md, with
the exception of C4.
In this case, the model \md\ provides a more precise value
than when using the seismic data alone, because the inclusion of 
atmospheric data helps to narrow down the range of possible values, when
the uncertainty in \mlsep\ is large.
The calculated solar value of 1,400 kg m$^{-3}$ is in excellent agreement
with the
true solar value (1,408 kg m$^{-3}$).

Comparing \logg$_{\nu}$ from Tables~\ref{tab:logg} and \ref{tab:loggmeand}
we find values that differ by at most 0.02 dex ($<1\sigma$).
This implies that by 
making a reasonable assumption about \teff, \logg\ can be 
well estimated using only \mlsep\ and \numax, or \numax\ alone.

Again, inspecting \logg\ from Tables~\ref{tab:logg} and \ref{tab:loggmeand}, 
but this time using the SEEK 
model-dependent values ('\logg$_{\rm MR}$')
we also find that, apart from C4, the derived \logg\ values are consistent
with and without the atmospheric constraints, but the uncertainties in the latter
reduce by a factor of 2 -- 3.
For C4 there is a difference of 0.1 dex between the two values, a change of 
nearly 2$\sigma$.  As noted before this discrepancy is due to the relatively 
larger uncertainty in \mlsep\ when no atmospheric constraints are available.

Comparing SEEK to the other pipeline results for \logg, 
we find $\sigma_{\rm sys}$ of
0.02, 0.05, 0.01, 0.04, 0.02, and 0.02 dex, respectively.
For the \kep\ stars, these differences are smaller than $\sigma_{\rm SEEK}$ 
except for C2.
The agreement between the \logg\ values from the five seismic
methods supports the results that we present, and 
we can be confident that any of the pipeline results using both
seismic and atmospheric parameters can determine \logg\ with 
a  minimal systematic bias, just as \citet{gai11} showed.

We also find that the VWA spectroscopic values of \logg\ agree 
with the model-dependent ones 
to within 1$\sigma$ for C1 and C2, 
and within 1.5$\sigma$ for C3 and C4.

\begin{table}
\begin{center}
\caption{Surface gravity and stellar mean density obtained 
from scaling relations ('$\nu$') and stellar models ('MR') 
using both seismic and
atmospheric constraints. 
 \label{tab:loggmeand}}
\begin{tabular}{ccccccccc}
\hline\hline
Star & \logg$_{\nu}$&\logg$_{\rm MR}$ & 
$\langle \rho\rangle_{\nu}$&
$\langle \rho\rangle_{\rm MR}$\\
&(dex)&(dex) & (kg m$^{-3}$) & (kg m$^{-3}$)\\
\hline
C1 & 3.86 \pmm\ 0.03& 3.88 \pmm\ 0.02  &175 \pmm\ 2 & 174 \pmm\ 4\\
C2 & 3.88 \pmm\ 0.03&3.88 \pmm\ 0.02  &185 \pmm\ 1 & 189 \pmm\ 2\\
C3 & 3.94 \pmm\ 0.03&3.97 \pmm\ 0.03  &254 \pmm\ 1 & 257 \pmm\ 5\\
C4 & 3.47 \pmm\ 0.03&3.52 \pmm\ 0.04 &39\pmm\ 16 & 46 \pmm\ 4\\
C5 & 3.79 \pmm\ 0.04&3.82 \pmm\ 0.03&135 \pmm\ 1 & 140 \pmm\ 2\\
Sun &...& 4.43 \pmm\ 0.01  &...& 1400 \pmm\ 18 & \\
\hline\hline
\end{tabular}
\end{center}
\end{table}

\subsection{Age}
\begin{figure}
\includegraphics[width=0.50\textwidth]{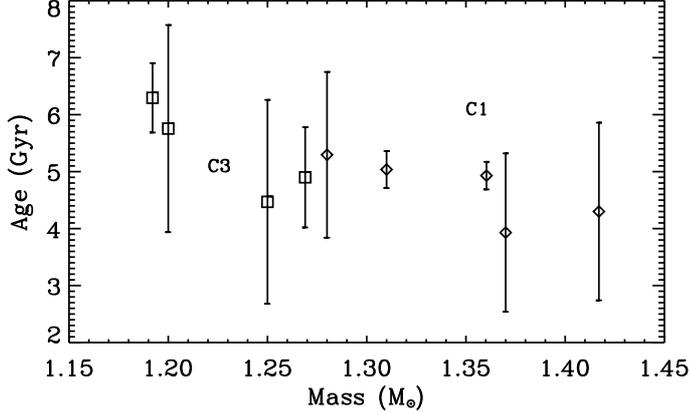}
\caption{Fitted age versus mass for stars C1 ($\diamond$) and C3 ($\square$)\label{fig:aage}}
\end{figure}

Determining the ages of the stars is a more complicated task since 
the value of the fitted age depends on the fitted mass and the
description of the model.
Figure~\ref{fig:aage} shows the fitted mass versus fitted age for 
stars C1 and C3 using five and four seismic methods, respectively 
(the CESAM2k/Mumbai method did not fit these data for C3 and C4).
For both stars a correlation between the two parameters can be seen;
a lower fitted mass will be matched to a higher age and vice versa.
However, the uncertainties from the SEEK method  ($\sim1$ Gyr for a 
mass of 1.37 \msol\ and $\sim2$ Gyr for a 1.25 \msol\ star) do 
capture the expected correlation with mass.

It is known that one of the physical ingredients to the models that
has a large effect on the age of the star is the convective core
overshoot parameter (see Sect.~\ref{sec:inputphysics} below) where 
fuel is replenished by mixing processes that in effect extends 
the lifetime of the star. 
The CESAM2k/Mumbai method includes several values for this parameter among
its grid of models, and this method fits an age of 5.3 Gyr for C1 
(the SEEK
method fits 3.9 Gyr).
However, the fitted mass of the star is also lower than the SEEK one, 
and if we consider
the uncertainty arising from the correlation with the age, then 5.3 Gyr
is not outside of the expected range.
For C2 the fitted mass for CESAM2k/Mumbai is 1.20 \msol\ (SEEK = 1.26 \msol),
and the fitted age is 4.4 Gyr (SEEK = 3.7 \pmm\ 0.7 Gyr).
The difference between the fitted ages is 1$\sigma$, again
not showing any difference due to the adopted values of the convective
core overshoot parameter.

The final value of the age depends on the adopted model (stellar properties
and physical description), but what seems to emerge from these data is that
all of the stars are approaching or have reached the end of the 
hydrogen burning phase.
In Table~\ref{tab:finalpar} we summarise the stellar properties obtained
by adopting \mlsep\ and \numax\ from Table~\ref{tab:seismicdata},  
the VWA spectroscopic constraints from Table~\ref{tab:finalpar}, and 
the photometric \teff\ for C5,  while
using the SEEK method (left columns) 
and combining the results from four (C3, C4, C5, Sun) and five (C1, C2)
grid-based methods (right columns).
We see that the mean values of the combined seismic methods results
in stellar properties consistent within 1.5$\sigma$ of the SEEK results.

\begin{table*}
\begin{center}
\caption{Stellar properties obtained   
using \mlsep, \numax, and the 
atmospheric constraints from VWA for C1 -- C4 and P/A for C5,
from SEEK (first half of table) and by combining the results from
all of the pipelines (second half of table).
\label{tab:finalpar}}
\begin{tabular}{lllllllllllllllll}
\hline\hline
Star&\logg&\rad & \mass&\age&$\langle \log g \rangle$&$\langle R\, \rangle$ & $\langle M\, \rangle$\\
&(dex)&(\rsol)&(\msol)&(Gyr) & (dex)&(\rsol) & (\msol)\\
\hline
C1& 3.88\pmm\ 0.02& 2.23\pmm\ 0.04& 1.37\pmm\ 0.11&  3.9\pmm\  1.4& 3.87& 2.21& 1.35\\
C2& 3.88\pmm\ 0.02& 2.11\pmm\ 0.05& 1.26\pmm\ 0.10&  3.7\pmm\  0.7& 3.89& 2.19& 1.37\\
C3& 3.97\pmm\ 0.03& 1.90\pmm\ 0.05& 1.25\pmm\ 0.13&  4.5\pmm\  1.8& 3.97& 1.88& 1.23\\
C4& 3.52\pmm\ 0.04& 3.81\pmm\ 0.19& 1.79\pmm\ 0.12&  1.1\pmm\  0.2& 3.52& 3.95& 1.87\\
C5& 3.82\pmm\ 0.03& 2.44\pmm\ 0.14& 1.44\pmm\ 0.26&  2.6\pmm\  0.9& 3.82& 2.47& 1.50\\
Sun& 4.43\pmm\ 0.01& 0.99\pmm\ 0.03& 0.97\pmm\ 0.06&  9.2\pmm\  3.8& 4.43& 0.99& 0.97\\

\hline\hline
\end{tabular}
\end{center}
\end{table*}

\subsection{Including \mssep\ in the seismic analysis
\label{sec:mssep}}
With the individual oscillation frequencies available, the value of 
\mssep\ can be easily calculated.  
This observable can be particularly useful for 
determining the stellar age because it is determined 
by the sound-speed gradient in the core of the star, and as hydrogen is burned,
a positive sound-speed gradient builds up. 
Three of the pipeline codes presented here do not allow one 
to use this value because it relies on the calculation of oscillation
frequencies for every model in the grid, unlike \mlsep\ which 
can be derived from a scaling relation (c.f. Eq.\ref{eqn:dnu}).
We included \mssep\ as an extra constraint in SEEK 
to study the possible changes in the values and uncertainties
of the stellar properties.

In Table~\ref{tab:finalparmssep} we list \logg, \rad, \mass, \age, and 
the uncertainties without (first line for each star) and 
with (second line for each star) \mssep\ as an observational constraint.
The first line for each star shows the same values as those presented
in Table~\ref{tab:finalpar}, and we repeat it here to make 
the comparison easier for the reader.
For C1 we obtained a more precise determination of the age. 
For C2 and C3, no improvements in the parameters were found.
For C5, all of the parameter uncertainties reduce by a factor of three,
and the values of the parameters change by 1$\sigma$.
The usefulness of \mssep\ here can most probably be explained by the 
lack of a metallicity constraint on the models.
For the Sun we also found that the age improved from an estimate of
9.2 Gyr to 4.9 Gyr, very close to the accepted value.
This last result can be explained by the fact that the stellar
'observables' change on a much slower scale than for a more evolved star
and hence provide weaker constraints.

\begin{table}
\begin{center}
\caption{Comparison between fitted stellar properties using SEEK
without (top lines) and with (bottom lines) \mssep\ 
as a seismic constraint.
\label{tab:finalparmssep}}
\begin{tabular}{lllllllllllllllll}
\hline\hline
Star&\logg&\rad & \mass&\age&\\
&(dex)&(\rsol)&(\msol)&(Gyr) \\
\hline
C1& 3.88\pmm\ 0.02& 2.23\pmm\ 0.04& 1.37\pmm\ 0.11&  3.9\pmm\  1.4\\
& 3.87\pmm\ 0.02& 2.22\pmm\ 0.05& 1.34\pmm\ 0.11&  4.5\pmm\  0.5\\
\\
C2& 3.88\pmm\ 0.02& 2.11\pmm\ 0.05& 1.26\pmm\ 0.10&  3.7\pmm\  0.7\\
& 3.88\pmm\ 0.02& 2.11\pmm\ 0.05& 1.25\pmm\ 0.10&  3.7\pmm\  0.6\\
\\
C3& 3.97\pmm\ 0.03& 1.90\pmm\ 0.05& 1.25\pmm\ 0.13&  4.5\pmm\  1.8\\
& 3.97\pmm\ 0.02& 1.90\pmm\ 0.06& 1.23\pmm\ 0.11&  5.0\pmm\  1.9\\
\\
C5& 3.82\pmm\ 0.03& 2.44\pmm\ 0.14& 1.44\pmm\ 0.26&  2.6\pmm\  0.9\\
& 3.85\pmm\ 0.01& 2.56\pmm\ 0.05& 1.71\pmm\ 0.09&  1.6\pmm\  0.2\\
\\
Sun& 4.43\pmm\ 0.01& 0.99\pmm\ 0.03& 0.97\pmm\ 0.06&  9.2\pmm\  3.8\\
& 4.43\pmm\ 0.01& 1.01\pmm\ 0.03& 1.01\pmm\ 0.04&  4.9\pmm\  0.5\\
\hline\hline
\end{tabular}
\end{center}
\end{table}

\section{Sources of systematic errors\label{sec:sectsourceserrors}}

\subsection{Using different spectroscopic constraints}
We adopted the VWA spectroscopic constraints because their \logg\ values
were in best agreement with the asteroseismic ones given
in 
Table~\ref{tab:logg}, 
and later confirmed in Sect.~\ref{sec:loggrhomr}.
However, 
the VWA parameters may not be the optimal ones, and so 
we should investigate possible systematic errors arising from 
using the other atmospheric parameters from Table~\ref{tab:atmos}.
We repeated the analysis using \mlsep\ and \numax\
with each set of atmospheric constraints
to derive radius and mass.
In Table~\ref{tab:input} we give the six sets of observational data 
for star C1, and we label these S1 -- S6.
We also give the fitted radius and mass with uncertainties for 
each of these sets using the SEEK method.

Inspecting sets S1 -- S5 we see that the largest differences
between the fitted properties is 0.05 \rsol\ and 0.12 \msol ($\sim$1$\sigma$), 
using
sets S1 and S2. 
These differences can be attributed to both 
the input \teff\ and \feh\ 
each contributing about equal amounts. 
At the level of precision of the spectroscopic \logg, this value has very
little or no role to play in determining the radius and the mass, due
to the tight constraints from the seismic data.

The results using S6, however, change by more than 1$\sigma$.  
This is clearly due to lacking metallicity constraints, 
and because C1 is considerably more metal-rich than the Sun.
If the star had solar metallicity, the results would be comparable between
each set.
For example, for C3 which has solar metallicity 
we find a comparable radius using the photometric data (1.89 \rsol)
and the spectroscopic data (1.85 -- 1.93 \rsol).
For C2 which has sub-solar metallicity we find the opposite trend to C1;  
the fitted radius is 2.18 \rsol\ without a metallicity constraint, 
while adopting  
the lower metallicity values 
yields smaller radii of 2.11 and 2.13 \rsol.
This same trend is found using the results from the other pipelines.

While the absence of a metallicity constraint in general increases the
uncertainties e.g. from 0.05 to 0.11 \rsol\ for C1, with 
similar values found for C2 and C3, its absence will also 
bias the final fitted value of radius and hence mass, although
still within its given uncertainty. 
For C4 the fitted radius varies by about 2\% correlating directly with
the input \teff\ but the statistical 
uncertainty is 4 -- 5\%.  
In this case the absence of a metallicity measurement is unimportant,
due to the error in \mlsep\ being a factor of 10
larger than for the other stars.

\begin{table}
\caption{Input spectroscopic observations (analysed along with the seismic data)
for star C1 and the SEEK determination of mass and radius.
\label{tab:input}}
\begin{center}
\begin{tabular}{llllllllll}
\hline\hline
Set \# &  \teff\ & \logg\ & \feh\ & \rad&$\sigma_{\rm R}$&\mass&$\sigma_{\rm M}$ \\
&(K)&(dex)&(dex)&(\rsol)&&(\msol)\\
\hline
S1  & 5717 &3.96&+0.35&2.25& 0.03& 1.45& 0.10\\
S2 &5445 & 3.84&+0.13& 2.20& 0.05& 1.33& 0.10\\

S3 &5580 & 3.81 & +0.19& 2.23& 0.04& 1.37& 0.11\\

S4  &5650 &	4.10&+0.36 &2.25& 0.04& 1.44& 0.10\\

S5 &5700 &	4.10&+0.13 & 2.23& 0.06& 1.39& 0.13\\

S6 &5660 &...	 &...  &2.13& 0.11& 1.21& 0.20\\

\hline\hline
\end{tabular}
\end{center}
\end{table}

\subsection{Different descriptions of the input physics\label{sec:inputphysics}}
We reported the fitted properties (mass, radius, and age) and
uncertainties for stars C1 -- C5 in Table~\ref{tab:finalpar} using 
the reference method SEEK and the mean values obtained by combining
the results from the four or five pipelines.
We analysed the data using five grid-based seismic 
methods to test for the reliability of the results.
Because the five methods are based on different sets of
physics and input parameters, we can consider the scatter of the results
from these pipelines
as indicative of the systematic error
that we can expect.
Testing all of the possible sources of systematic errors
in the most correct manner 
would imply generating numerous new grids
with every combination of physics, stellar evolution codes, and parameters,
and using the same analysis method that SEEK uses.  
While this is beyond the scope of this paper, and 
beyond the idea behind using grid-based methods, we can, however,
investigate systematics by varying a few of the 
most important physical ingredients of the models, for example, the 
convective core overshoot parameter, and the inclusion/exclusion of diffusion.
We keep in mind that 
the SEEK method covers almost the full range of possible mass, age, initial 
metal fraction, and mixing-length parameter for these stars, 
and the other pipelines use
different combinations of EOS, opacities, nuclear reaction rates,
and diffusion of elements (for some codes), and these should cover most of 
the systematics that one would expect to find.

We begun with a stellar model which is described by one set
of stellar parameters {\bf P}.
SEEK provides central parameters defined by their distribution in the $\chi^2$
plane,
so we used the SEEK
parameters as a starting point to 
determine a single best-fitting model using 
the Levenberg-Marquardt minimisation algorithm, and 
the same input physics as SEEK.
Because we have only five observations 
(\teff, \logg, \feh, \mlsep, \numax) and in principle 
five fitting parameters (\mass, \age, $Z_{\rm i}$, $X_{\rm i}$ or $Y_{\rm i}$, $\alpha$),
we decided to fix the initial hydrogen mass fraction, which in effect allows 
$Y_{\rm i}$ to vary slightly, such that $Y_{\rm i}$ has near solar value,
$Y_i = 0.273 - 0.278$ \citep{sb10}.
Table~\ref{tab:localmin} lists the values of the best fitted parameters
using the original input physics in SEEK 
and then for cases 1a -- 1d and 2a -- 2d, which are described below.

Once we have found {\bf P} we change the physical description
of the model by including convective core overshoot and 
setting its parameter to $\alpha_{\rm ov} = 0.25$, 
and we search again for a new
set of parameters that describe the observations best. 
We do this in various ways using the same minimisation algorithm.
The first time we fit the same four parameters (case 1a), the second time
we fit \age, $Z_{\rm i}$ and $\alpha$ (case 1b), 
then we fit \mass, $Z_{\rm i}$ and $\alpha$ (case 1c),
and finally we fit the parameters \mass\ and $\alpha$ only (case 1d).
If we fit both the mass and age together (case 1a), 
then we will determine a good
model, but it is not possible to test if the new fitted mass and age
are due to the 'correlation' term between the parameters, or due to
the new description of the model.  
For this reason we fix the mass and search for a set of parameters including
the age that 
adequately fit the observations (case 1b) and vice versa (case 1c).
This yields an estimate of the systematic error 
on the age/mass parameter for a fixed mass/age.
We also include case (case 1d) to eliminate the $M$--$Y_{\rm i}$ correlation e.g. 
\citet{met09,oze11}.
We repeat the same exercise while also including He diffusion as described
by Michaud \& Proffit (1993).  These cases are denoted by 2a -- 2d.

\begin{table*}
\begin{center}
\caption{The best-fitting parameters for star C1 found by using
a minimisation algorithm with SEEK physics with
some changes in the model.
\label{tab:localmin}}
\begin{tabular}{lllllllll}
\hline\hline
Description & $M$ & $R$ & $\tau$ & $Y_{\rm i}$ & $Z_{\rm i}$ & $\alpha$ & $\chi^2_R$\\
& (\msol)&(\rsol)&(Gyr)&&\\
\hline
& 1.359 &2.253 &3.93 & 0.2706 & 0.0294 & 1.21 & 1.26\\
(1a) $\alpha_{\rm ov} = 0.25$& 1.386&2.275 &3.81&0.2745&  0.0255&  1.33 & 1.73\\
(1b) $\alpha_{\rm ov} = 0.25$& ...&2.248& 3.70&0.2774&0.0226& 1.28& 1.89\\
(1c) $\alpha_{\rm ov} = 0.25$& 1.386&2.264& ... &  0.2739& 0.0261& 1.39& 1.49\\
(1d) $\alpha_{\rm ov} = 0.25$& 1.417 &2.290& ...& ... & ... &1.35 &  1.18\\
(2a) $\alpha_{\rm ov} = 0.25$, He diff& 1.379&2.265&3.74&0.2746& 0.0254& 1.30&1.75\\
(2b) $\alpha_{\rm ov} = 0.25$, He diff&...&2.250&3.55&0.2763&0.0237& 1.18&0.98\\
(2c) $\alpha_{\rm ov} = 0.25$, He diff&1.379&2.271&...&0.2745&0.0255& 1.40&1.22\\
(2d) $\alpha_{\rm ov} = 0.25$, He diff&1.429&2.295&...&...&...& 1.53&0.80\\
\hline\hline
\end{tabular}
\end{center}
\end{table*}

Inspecting Table~\ref{tab:localmin}, 
we find a maximum
difference of 5\% in mass and 1.8\% in radius (case 2d), 
while the largest difference in the fitted age is found for case (2b) 
resulting in a 10\% difference from the original fitted value.
These systematic errors are much smaller than the uncertainties 
given by SEEK in 
Table~\ref{tab:finalpar}.
While we do not claim that these values are typical values for all 
combinations
of physical descriptions in the models, these results indicate
that the uncertainties are realistic.
We may find larger differences using the more evolved
stars, but the uncertainties for these stars are also generally larger.

\section{Comparison of five grid-based approaches\label{sec:compare}}
Using the five \kep\ stars and the Sun as test cases for grid-based 
analyses we have
shown the following:
\subsection{Surface gravity}  
If the seismic data alone are used to determine \logg\ using stellar models,
then we can expect to find systematic differences between 
the pipelines due to the different combinations of input physics.
The difference between all of the results is at most 0.07 dex, which still 
provides a very strong constraint on \logg.
Combining the results from all of the pipelines yields \logg\ values 
in better agreement with those calculated directly from the seismic data 
(without models) when we make a modest assumption about \teff.
When both atmospheric and seismic constraints are
available, all of the pipelines yield consistent results for \logg, 
fitting to within at most 2$\sigma$ (or 0.05 dex) of either the SEEK result,
the mean value from all the pipelines, 
or the model-independent value. 
\subsection{Radius and mass}
The determination of mass, \mass, depends mostly on the value of the fitted radius, \rad,
because the seismic data constrains the $M-R$ correlation.
For this reason, similar trends are found for both \rad\ and \mass.
The systematic error ($\sigma_{\rm sys}$), defined here as the maximum difference
in fitted values between SEEK and the other pipelines, 
was found to agree within $\sim$2$\sigma$. 
We also found that in general for the radius $\sigma_{\rm sys}$ increases with
the evolutionary state of the star, i.e. 2\% for the Sun which is mid-MS, and 
10\% for the most evolved star in this study (C4).

We adopted SEEK as the reference method, which has been validated with
independent measurements of nearby stars, and we find
that for each star, the mean parameter value obtained by combining the 
results from all of the pipelines agrees to within 1$\sigma$ of 
the SEEK results, 
e.g. $R_{\rm SEEK} - \langle R\, \rangle < 1\sigma$ for all stars.
This seems to imply that adopting an average value of the stellar parameters
from several grid-based methods may be the optimal value to use.  
Nevertheless, all of the grids provide radius and mass values consistent
within 2.4$\sigma$.

\subsection{Age} 
In Fig.~\ref{fig:aage} we showed the fitted mass versus age 
for stars C1 and C3 using all
of the methods for C1 and four methods for C3. While the returned value
of the age depends upon the description of the physics and the mass of the
star, in general we found the differences among the derived age is due to 
the mass-age correlation.  Both the RADIUS and Yale-Birmingham pipelines provide
the smallest uncertainties for C1. These are probably underestimated if 
one is to consider all possible sources of systematic errors.

When \mssep\ is available it is optimal to 
use the methods 
based on oscillation frequencies (SEEK, CESAM2k/Mumbai) in
order to decrease the uncertainty in the age, especially 
for the less evolved MS stars e.g. for the Sun we find 9.2 \pmm\ 3.8 Gyr
without \mssep\ and 4.9 \pmm\ 0.5 Gyr including it.
However, the actual fitted value of age does not change very much for 
the more evolved MS stars analysed here.

\section{Additional stellar properties from complementary data\label{sec:indep}}

\subsection{Determination of distance\label{sec:dist}}

\begin{table}
\begin{center}\caption{Photometric data for C1 -- C5 from the literature.\label{tab:vebv}}
\begin{tabular}{llllllll}
\hline\hline
Star & $V^{a}$  & $V^{b}$ & $V^{c}$ & E($B-V$)$^{d}$\\ 
& (mag) & (mag) & (mag) & (mag)\\
\hline\hline
C1 & ... & 10.942 & 10.901 & 0.028 \\
C2 & 10.871 & 10.840 & 10.959 & 0.056\\
C3 &  11.684 & 11.635 & 12.004 & 0.070\\ 
C4 &  11.982 & 11.885 & 12.160 &  0.075\\
C5 &   ... & ... & 12.049 & 0.062\\
\hline\hline
\end{tabular}
\end{center}
\begin{tiny}$^{a}$ \citet{urb98}
$^{b}$ \citet{kha01}
$^{c}$ \citet{dro06}
$^d$ KIC
\end{tiny}
\end{table}

The luminosities corresponding to the models from Table~\ref{tab:finalpar} yield values of the absolute bolometric magnitude $M_{\rm bol}$.  
To obtain the absolute $V$ magnitude $M_V$ we interpolate the tables from \citet{flo96} for the values of \teff\ given by VWA in Table~\ref{tab:atmos}
(and the photometric value for C5)
to obtain the bolometric correction BC, and apply this 
correction to $M_{\rm bol}$.
In Table~\ref{tab:vebv} we list the apparent $V$ magnitudes from various sources.
To account for reddening 
we use the standard extinction law A$_{\rm V}$ = 3.1$\times$E($B-V$), 
where 3.1 is a typical value \citep{sm79} and 
E($B-V$) are obtained from the KIC.
We note that the adopted E($B-V$) should be used with
caution \citep{mol09}.
The distance modulus is calculated from $M_V$ and 
the de-reddened $V$, to give the 
asteroseismic distance $d$ in parsecs using $d = 10^{0.2(V-M_V)+1}$.
In Table~\ref{tab:dist} we list the model \lum, $M_V$, the de-reddened $V$ magnitude adopting the values from \citet{dro06} $V_{\rm der}$, and the distance $d_{06}$, with   
the uncertainties arising from the luminosity uncertainty. 
The column with heading $d_{01}$ shows the distance using $V$ from \cite{kha01}.
We are able to determine distances to these stars with a 
precision of less than 10\%.
However, for C2 and C4 we find a discrepancy of 14\% and 12\% due 
to the different reported magnitudes.

We can also calculate the distance using one of the surface brightness
relations from \citet{ker04} (Table 5), if we know the radius. 
We used the relationship for \teff\ and $V$; 
$\log \theta = 3.0415 (\log T_{\rm eff})^2 -25.4696 \log T_{\rm eff}+53.7010-0.2V_{\rm der}$, where $\theta$ is the angular diameter in milliarcseconds.  
Using the \citet{dro06} magnitudes, the fitted radii, and 
the observed \teff, 
we calculated the distance according to
this relation $d_{\rm SB}$ (also given in Table~\ref{tab:dist}),  and 
these were found to agree with those using the distance modulus.

\begin{table*}
\begin{center}\caption{Model luminosities \lum, absolute $V$ magnitudes $M_V$, 
de-reddened magnitudes $V_{\rm der}$, and asteroseismic distances from the 
standard distance modulus equation ($d_{06,01}$) and surface brightness 
relation $d_{\rm SB}$.
\label{tab:dist}}
\begin{tabular}{lllllllllll}
\hline\hline
Star &\lum ($\sigma_{\rm L}$) & $M_V$ & $V_{\rm der}$  & $d_{06}$ &$d_{01}$&$d_{\rm SB}$\\
& (\lsol) & (mag) & (mag) &(pc)&(pc)&(pc) \\
\hline\hline
C1 &  4.2 (1.1)&3.300&10.814& 318$^{+39}_{-45}$& 324$^{+40}_{-46}$&318\pmm\   20\\
C2 &  5.3 (1.1)&2.967&10.785& 366$^{+36}_{-40}$& 347$^{+34}_{-38}$&371\pmm\   24\\
C3 &  3.6 (1.2)&3.427&11.787& 470$^{+73}_{-86}$& 396$^{+61}_{-73}$&475\pmm\   31\\
C4 & 20.0 (1.1)&1.510&11.927&1212$^{+33}_{-34}$&1068$^{+29}_{-30}$&1231\pmm\   95\\
C5 &  7.8 (1.1)&2.526&11.857& 735$^{+50}_{-54}$&...&741\pmm\   61\\
\hline\hline
\end{tabular}
\end{center}
\end{table*}

\subsection{Rotational period and inclination \label{sec:pi}}
The value of \vsini\ can be determined from the spectroscopic analysis (see Table~\ref{tab:atmos}). 
Combining this with \rad\ 
allows us to constrain the stellar rotational period $P_{\rm ROT}$ -- inclination $i$ relation.
We use the \vsini\ values from ROTFIT to determine this relationship.
With no observational constraints on $i$ for any star and having 
uncertainties in \vsini\ of the same order as its value for C3 (implying possibly \vsini\ $\sim0$ \kms),
the lower bound on $P_{\rm ROT}$ is unconstrained for all of the stars, while
the upper bound is poorly restricted for C3.  
We can place an upper bound on $P_{\rm ROT}$ for C1, C2, and C4 of 
384, 64, and 168 days, respectively.

Both \citet{cam11} and \citet{mat11} 
analysed the low range of the frequency spectrum
to look for signatures of a rotation period.  
They estimate $P_{\rm ROT}$ for C1, C2, and C3 of $\sim$ 36, 23, and 27 
days, respectively, consistent with our results. 
Adopting these values as $P_{\rm ROT}$ constrains 
the stellar inclination angle  
$i = 20^{\circ +18}_{~-15}$ for C1,
$i = 44^{\circ +46}_{~-23}$ for C2, and  
$i = 25^{\circ +65}_{~-20}$ for C3.

\subsection{Lithium content and age estimate\label{sec:li}}

\begin{figure}
\centering
\includegraphics[width=0.5\textwidth]{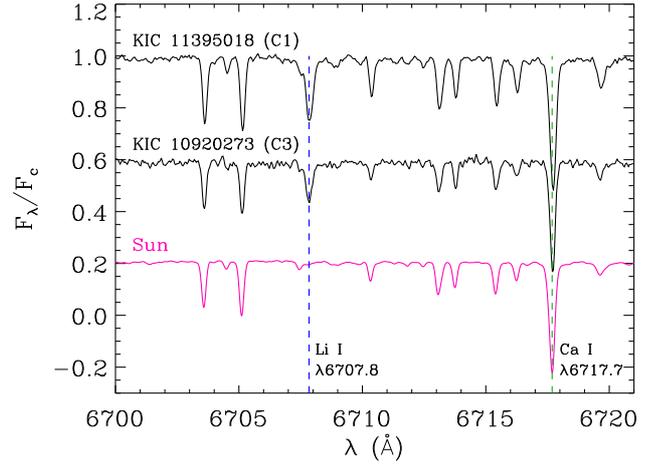}
\caption{A region of the optical spectra which contains the 
Lithium $\lambda$6707.8\,\AA\ line.
From top to bottom we show the spectra from C1, C3 and 
a high-resolution solar 
spectrum (Ganymede taken in 2007 with HARPS).} 
\label{fig:lithium}
\end{figure}

We noted that two of the investigated stars, namely C1 and C3, clearly display an 
Li\,{\sc i}\,$\lambda$6707.8\,\AA\  photospheric absorption line in the FIES spectra (Fig.~\ref{fig:lithium}),
while for the remaining two stars the lithium line is not detectable.
Lithium is burned at relatively low temperatures in stellar interiors ($\sim2.5\,\times\,10^6$\,K).
As a consequence, it is progressively depleted from the stellar atmospheres of late-type stars 
when mixing mechanisms pull it deep into their convective layers. 
Therefore, its abundance can be used for estimating the stellar age, as shown by \citet{Sestito05} for
stars belonging to 22 open clusters spanning the age range 0.005--8 Gyr.

We measured the equivalent width of the lithium line, correcting for the small contribution of the nearby 
Fe-I\,$\lambda$6707.4\,\AA\  line as suggested by \citet{sod93}, and found 
$W_{\rm Li}=102\pm10$\,m\AA\  and $W_{\rm Li}=55\pm10$\,m\AA\ 
for C1 and C3, respectively.
For C2 and C4, we estimated an upper limit for the lithium equivalent width $W_{\rm Li}< 10$\,m\AA\ 
as the product of the error of the normalised flux 
per spectral point and the integration width 
($\simeq 1$\,\AA).

We derived a lithium abundance $\log n({\rm Li})= 2.6\pm0.1$ and $\log n({\rm Li})= 2.4\pm0.1$
for C1 and C3 by interpolation of the NLTE curves of growth tabulated by \citet{PavMag96}, where by definition $\log n({\rm H}) = 12$.
For C2 and C4, the lithium abundance is $\log n({\rm Li}) < 1.9$.
The correlation of lithium abundance and age established by \citet[][see their Fig.~7]{Sestito05}, 
suggests an age of 
0.1 -- 0.4\,Gyr and 1 -- 3\,Gyr for C1 and C3, respectively, 
while their Table~3 suggests an age for C2 and C4 
corresponding to the most evolved clusters 
in their study (M67 at 5 Gyr).
The asteroseismic ages of 3 Gyr for C2 and 1 Gyr for C4 imply at 
least evolved MS stars for masses of 1.4 and 1.9 \msol, respectively.

New $W_{\rm Li}$ measurements for solar-like stars in galactic open clusters in the age range 1--8\,Gyr, 
have 
shown that the low solar lithium abundance is not the standard for a star of that age
and mass \citep{Randich10}. 
In particular, \citet{Pasquini08}  found a spread of lithium abundance in stars of the 
 solar-age cluster M\,67, ranging from $\sim$0.4 to $\sim$2.0 dex. 
 \citet{Randich10} also found that 
the average abundance for various clusters of similar age varies. 
 For stars in the $T_{\rm eff}$ range 5750--6050\,K,
some evidence of bimodality in lithium depletion after 1\,Gyr seems to emerge from these new data, with some clusters following the
solar behaviour of abundance decay and other ones forming a plateau with $\log n({\rm Li})= 2.2-2.4$. 
It is now unclear what the driving mechanism for depletion is,
 but it appears that some other 
parameter apart from mass and age is playing a role. 

This could reconcile the lithium abundance  $\log n({\rm Li})= 2.4\pm0.1$ of C3 with the higher age 
derived by asteroseismology. However, for C1, whose temperature is significantly lower (more lithium depletion),
the lithium abundance of $\log n({\rm Li})= 2.6\pm0.1$ is not compatible with the age of 4 Gyr deduced by asteroseismology.
We note that C1 is the most metal-rich star studied in this paper.

\section{Summary and Conclusions}

In this work we analysed the five solar-type stars KIC~11395018, KIC~10273246, 
KIC~10920273, KIC~10339342, and KIC~11234888, referred to as C1 -- C5, respectively, 
based on more than eight months of short-cadence \kep\ data.
The global seismic quantities (\mlsep\ and \numax) coupled with
atmospheric parameters derived from photometry and spectra taken with the NOT
telescope yield
stellar properties  
 with an average precision of 2\% in mean density, 
$\sim$0.03 dex in surface gravity,
 2--5\% in radius,
7--11\% in mass,
and 35\% in age (Table~\ref{tab:end}).

We used five grid-based approaches based on stellar models 
to estimate systematic errors,
since the grids are all based on different physics and evolution codes.
We found an agreement between all of the methods
to within 2.4$\sigma$ for mass and radius, and these values were
often comparable to the statistical uncertainties
(see Tables~\ref{tab:radius} and \ref{tab:amass}).
However, we also tested some specific sources of systematic errors arising
from including convective core overshoot and settling of helium.
We found that the fitted stellar parameters (radius, mass, and age)
changed by an amount
smaller than the 
statistical uncertainties reported in Table~\ref{tab:end}.
These changes were also smaller than the dispersion among the results from
the five grid-based methods, and we can therefore conclude
 that using several different methods provides
an estimate of the systematic errors. 
However, since the value of the age is highly model-dependent, and we only
tested two physical ingredients in the models, 
the systematic errors in the age could be larger.
In Table~\ref{tab:end2} we summarise the estimates of the 
systematic errors $\sigma_{\rm sys}$ from various sources.

The spectroscopic data were analysed by five groups independently, 
each providing atmospheric parameters that varied by more than 200 K in \teff\ and 0.2 dex for \feh.
The differences obtained can be attributed to the 
quality of the spectroscopic data (low S/N and intermediate resolution).
However,  we found that the results are in general correlated
(Fig.~\ref{tab:atmos}), indicating
that with higher resolution and higher S/N data, the results should be in much
better agreement.   
The different results obtained, however, also
allowed us to study the influence of these atmospheric
parameters for determining 
the radius and mass of the stars.
We found that the fitted radius depends on both the adopted \teff\ and \feh, 
and the different values led to a discrepancy of about 2.5\% in 
stellar radius and 
8\% in stellar mass. 
In the absence of an \feh\ measurement we found biased results in the fitted
radius and mass for stars with non-solar metallicity 
(metal-rich C1 and metal-poor C2), and larger uncertainties.
Higher-quality spectra should be obtained for these targets in the near future to match the exceedingly high quality of the \kep\ data.

We investigated the role of including \mssep\ as 
an observational constraint with \mlsep\ and \numax\ 
(Sect.~\ref{sec:mssep} and Table~\ref{tab:finalparmssep}), and 
we found that 
\mssep\ is important primarily for the uncertainty in age for C1, C5,  
and the Sun. For C5 this is due to the absence of an \feh\ measurement.
We suspect that its larger impact for the Sun, 
is because it is in the middle of its MS lifetime
where the other atmospheric observables change relatively
slowly with the age, thus providing weaker constraints.

Coupling photometric magnitudes with the model luminosities, we 
derived distances with a precision of less than 10\%,
and for 
C3 and C4, systematic errors of 14\% and 12\% respectively arise from the 
$V$ magnitudes reported by different authors.
Later on in the mission, parallaxes may be obtained for the unsaturated stars.
 These independent determinations of distances should help to resolve 
the differences in the $V$ magnitudes and/or reddening.

By coupling the derived radius with the observed \vsini, limits can be placed on
the rotational period $P_{\rm ROT}$ of the stars.  
For C1, C2, and C4 we can impose an upper bound on $P_{\rm ROT}$ of 386, 67, and 177 days, respectively.
Rotational period, inclination, and rotational velocity can all be determined in an independent manner
by studying either the stellar spot distribution using the time series data, the frequency splittings in the power spectrum, or by studying the low frequency range of the power spectrum.
\citet{cam11} and \citet{mat11} study the low frequency range of the power spectra and estimate 
rotational periods of 36, 23, and 27 days for C1, C2, and C3, respectively.
Adopting these as $P_{\rm ROT}$ constrains 
the stellar inclination angle  $i = 20^{\circ +18}_{~-15}$ for C1,
$i = 44^{\circ +46}_{~-23}$ for C2, and  
$i = 25^{\circ +65}_{~-20}$ for C3.

The amount of lithium absorption in the atmospheric spectra can 
be used to estimate the age of the star.  
For the stars C2 and C4 no Li absorption is seen, indicating  
evolved MS stars,  
and this is confirmed by the seismic analysis.
For C3, the Li abundance indicates an age 
that is inconsistent with the seismic value, but these could be reconciled, by 
considering new results which show a bimodal
distribution of Li depletion \citep{Randich10}.  
For the metal-rich star C1, 
however, the relatively high Li abundance indicates a non-evolved MS star, 
while the
asteroseismic analysis suggests an incompatible value of $\sim4$ Gyr.

\citet{cam11} and \citet{mat11} analysed the time series 
and presented the oscillation frequencies for C1, C2, C3, and C5.  
Using the individual frequencies will yield more precise stellar properties, as well as studies of the 
stellar interior (Brand\~ao et al. 2012, Do\u{g}an 
et al. 2012) as shown by \citet{met10}.
Up to now, such precise values 
have only been possible for members of detached 
eclipsing binaries, the Sun, and a few very bright stars.  
Thanks to the significant improvement in data quality from \kep, 
this is about to change. 

\begin{table*}
\caption{Summary of stellar properties for 
KIC~11395018 (C1), KIC~10273246 (C2), 
KIC~10920273 (C3), KIC~10339342 (C4),
and KIC~11234888 (C5), derived from atmospheric and 
mean seismic parameters, and stellar models.\label{tab:end}}
\begin{center}
\begin{tabular}{lcccccccccccccccccccccrrrrrrrrrrrcccccccc}
\hline\hline
& C1 & C2 & C3 & C4 & C5\\
\hline
$^a$\md$_{\nu}$ (kg m$^{-3}$)&175\pmm2&185\pmm1&254\pmm1&39\pmm16&135\pmm1\\
$^b\langle \rho \rangle_{\rm MR}$ (kg m$^{-3}$)&174\pmm4 &189\pmm 2&257\pmm 5&46\pmm4&140\pmm 2\\
$^a$\logg$_{\nu}$ (dex)&3.86\pmm0.03&3.88\pmm0.03&3.94\pmm0.03&3.47\pmm0.03&3.79\pmm0.04\\
$^b$\logg$_{\rm MR}$ (dex)&3.88\pmm0.02&3.88\pmm0.02&3.97\pmm0.03&3.52\pmm0.04&3.82\pmm0.03\\
\rad\ (\rsol)&2.23\pmm0.04&2.11\pmm0.05&1.90\pmm0.05&3.81\pmm0.19&2.44\pmm0.14\\
\mass\ (\msol)&1.37\pmm0.11&1.26\pmm0.10&1.25\pmm0.13&1.79\pmm0.12&1.44\pmm0.26\\
\age\ (Gyr)&3.9\pmm1.4&3.7\pmm0.7&4.5\pmm1.8&1.1\pmm0.2&2.6\pmm0.9\\
$^c$\age$_{\langle \delta \nu \rangle}$ (Gyr)&4.5\pmm0.5&3.7\pmm0.6&5.0\pmm1.9&...&1.6\pmm0.2\\
\lum\ (\lsol)&4.2\pmm1.1&5.3\pmm1.1&3.6\pmm1.2&20.0\pmm1.1&7.8\pmm1.1\\
\teff$_{\rm model}$ (K)&
5547&   6047&      5789&      6255&      6180
\\
$i$ ($^{\circ}$) &$20^{+18}_{-15}$&$44^{+46}_{-23}$&$25^{+65}_{-20}$&...&...\\
$P_{\rm ROT max}$ (days)&384 &  64 &...& 168&...\\
$^dP_{\rm ROT est}$ (days)&36 (6) &  23 &27&...&19--27 (5) \\
$d$ (pc)&318$^{+39}_{-44}$&366$^{+36}_{-40}$&470$^{+72}_{-86}$ & 1212$^{+33}_{-34}$&
735$^{+50}_{-54}$\\
\hline\hline
\end{tabular}
\end{center}
$^{a,b}$ Subscripts $\nu$ and MR indicate that the value was obtained
directly from the data and from the models, respectively.\\
$^c \tau_{\langle \delta \nu \rangle}$ is when \mssep\ is
included as an observational constraint.  In this case the uncertainties
in \logg, \rad, and \mass\ for C5 reduce by a factor of three.\\
$^d P_{\rm ROT est}$ as reported by \citet{cam11} and \citet{mat11}, with uncertainties in parenthesis.
\end{table*}

\begin{table*}
\caption{Estimates of systematic errors in the stellar properties, given in CGS units for $g$, solar units for \rad\ and \mass, and Gyr for age, with
\% values shown in parentheses. 
\label{tab:end2}}
\begin{center}
\begin{tabular}{llllllllllllllllllllllcccccccccccccccccccccrrrrrrrrrrrcccccccc}
\hline\hline
& C1 & C2 & C3 & C4 & C5\\
\hline
$\sigma_{\log g,{\rm grid}}$  &0.02&0.05&0.01&0.04&0.02\\
$\sigma_{R,{\rm grid}}$ & 0.06 (3)&0.22 (11) &0.07 (4) &0.38 (10) &0.12 (5)\\
$\sigma_{M,{\rm grid}}$ & 0.09 (7)&0.43 (34) &0.06 (5)&0.09 (5)&0.18 (12)\\
$\sigma_{\tau,{\rm grid}}$ & 1.4 (36) & 1.7 (38) & 1.8 (40)&0.2 (18) &0.4 (16)\\
$\sigma_{R,{\rm spec}}$ & 0.05 (2) &0.08 (4)& 0.03 (2) & 0.18 (5)& ...\\
$\sigma_{M,{\rm spec}}$ & 0.13 (9)&0.16(13) & 0.03 (2)&0.12 (7)& ...\\
$\sigma_{R,{\rm phys}}$ & 0.06 (3)&...&...&...&...\\
$\sigma_{M,{\rm phys}}$ & 0.05 (4)&...&...&...&...\\
$\sigma_{\tau,{\rm phys}}$ & 0.35 (9)& ...&...&...&...\\
\hline\hline
\end{tabular}
\end{center}
 The subscripts 'grid', 'spec', and 
'phys' mean estimates of systematic errors from using different grids/pipelines, different atmospheric constraints, and different descriptions of the 
physics in the models, respectively 
(see Sects.~\ref{sec:sectstellarproperties} and \ref{sec:sectsourceserrors} 
for details).
\end{table*}

\begin{acknowledgements}
All of the authors acknowledge the \kep\ team for their years of work to
provide excellent data.
Funding for this Discovery mission is provided by NASA's Science
Mission Directorate.
This article is based on observations made with the Nordic Optical
Telescope operated on the island of La Palma 
in the Spanish Observatorio del Roque de los Muchachos. 
We thank Othman Benomar and Fr\'ed\'eric Thevenin for useful discussions, and
we also thank the referee for very constructive comments which has 
greatly improved
 the manuscript.
Part of this research was carried out while OLC was
a Henri Poincar\'e Fellow at the Observatoire de la C\^ote d'Azur. 
The
Henri Poincar\'e Fellowship is funded the Conseil G\'en\'eral des
Alpes-Maritimes and the Observatoire de la C\^ote d'Azur.
DSt acknowledges support from the Australian Research Council.
NG acknowledges the China State Scholarship Fund that allowed her to
spend a year at Yale. She also acknowledges grant 2007CB815406 of the
Ministry of Science and Technology of the Peoples Republic of China
and grants 10773003 and 10933002 from the National Natural Science
Foundation of China. WJC and YE acknowledge the
financial support of the UK Science and Technology Facilities Council
(STFC), and the International Space Science Institute (ISSI).
IMB is supported by the grant SFRH / BD / 41213 /2007 funded
by FCT / MCTES, Portugal.
EN acknowledges financial support of the NN203 302635 grant from the
MNiSW.
GD, HB, and CK acknowledge financial support from The Danish Council for Independent Research and thank Frank Grundahl and Thomas Amby Ottosen for suggestions regarding the NOT proposal.
JB, AM, AS, and TS acknowledge support from the National Initiative on
Undergraduate Science (NIUS) undertaken by the Homi Bhabha Centre for
Science Education -- Tata Institute of Fundamental Research (HBCSE-TIFR),
Mumbai, India.
SGS acknowledges the support from grant SFRH/BPD/47611/2008
from the Funda\c{c}\~ao para a Ci\^encia e Tecnologia (Portugal).
JM\.Z acknowledges the Polish Ministry grant number N N203 405139.
DSa acknowledges funding by the Spanish Ministry of Science and Innovation
(MICINN) under the grant AYA 2010-20982-C02-02.

\end{acknowledgements}

\appendix

\section{Oscillation frequencies}
In Table~\ref{tab:ff} we list the published oscillation frequencies 
from \citet{cam11} and \citet{mat11} for stars C1, C2, C3, and C5.
\begin{table*}
\begin{center}
\caption{Published frequencies given in \mhz\ for stars  
KIC~11395018 (C1),
KIC~10273246 (C2), 
KIC~10920273 (C3), and 
KIC~11234888 (C5), as reported by \citet{cam11} and \citet{mat11}.\label{tab:ff}}
\begin{tabular}{lllllllllllllllllllllll}
\hline\hline
 $l$ & KIC~11395018 & KIC~10273246 &KIC~10920273 & KIC~11234888\\
& C1 & C2 & C3 & C5\\
\hline
0&686.66 \pmm\ 0.32 & 737.90 \pmm\ 0.30&  826.66$^a$ \pmm\ 0.25   & 627.67 \pmm\ 0.19   \\
0&732.37 \pmm\ 0.18 & 785.40 \pmm\ 0.20&  882.77 \pmm\ 0.20       & 669.35 \pmm\ 0.16 \\
0&779.54 \pmm\ 0.14 & 833.90 \pmm\ 0.20&  939.58  \pmm\ 0.16      & 711.63 \pmm\ 0.15   \\
0&827.55 \pmm\ 0.15 & 883.50 \pmm\ 0.20&  997.14  \pmm\ 0.18      & 753.64 \pmm\ 0.20 \\
0&875.40 \pmm\ 0.16 & 932.70 \pmm\ 0.50&  1054.33 \pmm\ 0.30      & 794.56 \pmm\ 0.20  \\
0&923.16 \pmm\ 0.19 & 981.10 \pmm\ 0.30&  1111.51 \pmm\ 0.25      & 836.83 \pmm\ 0.21  \\
0&971.05 \pmm\ 0.28 & 1030.70 \pmm\ 0.40& 1170.77$^a$ \pmm\ 0.33  & 877.80$^a$ \pmm\ 0.22  \\
0&  ...                & 1079.30 \pmm\ 0.20& 1226.34$^a$ \pmm\ 0.33  &   ...   \\
\\
1&667.05$^a$ \pmm\ 0.22 & 622.80 \pmm\ 0.20     & 794.65$^{b}$ \pmm\ 0.32  & 506.72  \pmm\  0.20    \\
1&707.66 \pmm\ 0.19     &661.90 \pmm\ 0.50     & 838.61$^{b}$ \pmm\ 0.25  &  563.30 \pmm\ 0.14  \\
1&740.29$^b$ \pmm\ 0.17 & 695.75$^b$ \pmm\ 0.27 & 914.52 \pmm\ 0.16        &  594.83 \pmm\ 0.16  \\
1&763.99 \pmm\ 0.18     & 724.70 \pmm\ 0.20     & 968.19 \pmm\ 0.13        &  686.34 \pmm\ 0.17  \\
1&805.74 \pmm\ 0.13     & 764.30 \pmm\ 0.30     & 1023.58 \pmm\ 0.14       &  741.10 \pmm\ 0.18  \\
1&851.37 \pmm\ 0.11     & 809.80 \pmm\ 0.20     & 1079.10 \pmm\ 0.31       &  815.43 \pmm\ 0.21  \\
1&897.50 \pmm\ 0.15     &  857.30 \pmm\ 0.20     & 1135.36$^c$ \pmm\ 0.31   &  855.67 \pmm\ 0.21  \\
1&940.50 \pmm\ 0.15     & 905.60 \pmm\ 0.30& ...  & ...     \\
1&997.91 \pmm\ 0.33     &950.00 \pmm\ 0.30&  ...  & ...     \\
1&   ...                      & 1008.60 \pmm\ 0.40&  ...  & ...    \\
1&        ...                  &  1056.30 \pmm\ 0.20& ... & ...      \\
1&     ...                     &1103.30 \pmm\ 0.40&  ...  & ...    \\
\\
2 & 631.19$^a$ \pmm\ 1.36 & 688.50 \pmm\ 0.70    & 822.39$^a$ \pmm\ 0.28 & 582.84$^a$  \pmm\ 0.21  \\
2&680.88 \pmm\ 0.45      & 734.80 \pmm\ 0.60   & 873.10$^{a,d}$ \pmm\ 0.32& 624.65 \pmm\ 0.18  \\
2&727.78 \pmm\ 0.30      &  779.50 \pmm\ 0.40   & 934.49 \pmm\ 0.22&  708.67 \pmm\ 0.19  \\
2&774.92\pmm\ 0.16       & 830.30 \pmm\ 0.40    & 992.44 \pmm\ 0.13&  751.84  \pmm\ 0.21  \\
2&823.50 \pmm\ 0.16      &  880.60 \pmm\ 0.50   & 1049.36 \pmm\ 0.39& ... \\
2&871.29 \pmm\ 0.21       &  927.50 \pmm\ 0.40   & 1106.76 \pmm\ 0.34 & ...  \\
2&918.10 \pmm\ 0.28       &  977.60 \pmm\ 0.40   &  ...& ...  \\
2&965.83 \pmm\ 0.23       &1025.30 \pmm\ 1.30& ... & ...  \\
2 & 1016.61$^{a,f}$ \pmm\ 0.73 &1073.70 \pmm\ 0.20& ...& ...   \\
2 &   ...                     &1122.70$^{a,d}$ \pmm\ 0.40 &  ...& ...  \\
\\
\hline\hline
\end{tabular}
\end{center}
The frequencies were fitted with 11, 10, 10, and 11 different methods, and a criteria was imposed to obtain
a minimal frequency set representative of the results from most of the fitters.\\
$^a$ Frequencies not among the minimal set. \\
$^b$ $l=1$ mixed mode.\\ 
$^c$ Mode close to the second harmonic of the inverse of the long-cadence period.\\
$^d$ Possible $l=2$ mixed mode introduced a posteriori.\\
$^e$ Uncertain identification.\\
$^f$ Mode tagging changed a posteriori from $l=0$ to $l=2$.\\

\end{table*}

\section{Atmospheric analysis methods\label{appa}}

\subsection{BIA}
The MOOG code (version 2002; \citealt{sne73}) determines the iron abundance under the 
assumption of local thermodynamic equilibrium (LTE), using a grid of 1D model atmospheres by 
\cite{kur93}. The LTE iron abundance of C1 -- C4 was derived from the equivalent widths of 55--68 Fe-I 
and 11 Fe-II lines in the 4830--6810 \AA~range, measured with a Gaussian fitting procedure in 
the IRAF\footnote{IRAF is distributed by the National Optical Astronomy Observatory, which is operated 
by the Association of the Universities for Research in Astronomy, inc. (AURA) under cooperative 
agreement with the National Science Foundation.} task splot. For the analysis, we used the same list of 
lines as \cite{bia11} and we followed their prescriptions. In particular, \teff\ and \vturb\
were determined by requiring that the iron abundance be independent of the excitation potentials 
and the equivalent widths of Fe-I lines. The surface gravity was determined by requiring ionization 
equilibrium between Fe-I and Fe-II. The initial values for \teff, \logg\, and \vturb\
were chosen to be solar ($T_{\rm eff}$ = 5770 K, $\log g$ = 4.44 dex, 
and $\xi_{\rm t}$ = 1.1 km s$^{-1}$).

\subsection{ROTFIT}

The IDL\footnote{IDL (Interactive Data Language) is a registered trademark of ITT Visual Information 
Solutions.} code ROTFIT \citep{fra06} performs a simultaneous determination of $T_{\rm eff}, 
\log g$, and [Fe/H] for a star, as well as its projected rotational velocity $v\sin i$, by comparing 
the observed spectrum with a library of spectra of reference stars (see, e.g., \citealt{kat98,sou98}). 
We estimated the stellar parameters and $v\sin i$ as the
mean values for the 10 reference stars whose spectra, artificially
broadened with $v\sin i$ in steps of 0.5\,km\,s$^{-1}$, most
closely resembled the target spectrum, and each is quantified by  a $\chi^2$ 
measure. 
We adopted their standard deviation as a measure of the uncertainty. 
We applied the ROTFIT code to all echelle orders which cover the 
range 4300--6680 \AA\  in the observed spectrum. The adopted estimates for
the stellar parameters come from a weighted mean of the values for all the individual orders,
where more weight is assigned to the best fitted or higher S/N orders,
and the ``amount of information" contained in each spectral region expressed by the total spectral-line absorption.
The standard error of the weighted mean was adopted as the uncertainty estimate for the final values of 
the stellar parameters. 
With the same code, we also performed an MK
spectral classification for our targets.

\subsection{NIEM}
The NIEM analysis follows the methodology presented in 
\citet{niem09}
and relies on an efficient spectral synthesis based on a least
squares optimisation algorithm (see  \citealt{tak95}). This method allows for
the simultaneous determination of various parameters involved with
stellar spectra and consists of the minimisation of the deviation
between the theoretical flux distribution and the observed normalised
one. The synthetic spectrum depends on the stellar parameters: 
effective temperature \teff, surface gravity \logg, microturbulence \vturb,
rotational velocity \vsini, radial velocity \vrad, and the relative
abundances of the elements. 
The first three parameters are obtained before the
determination of abundances of chemical elements and they are
considered as input parameters. All the other above-mentioned parameters
can be determined simultaneously because they produce detectable and
different spectral signatures. 

\teff, \logg, and \vturb\ are adjusted by comparing the abundances
determined from the unblended 
Fe-I and Fe-II lines. 
We require that the abundances
measured from Fe-I and Fe-II 
yield the same result. The absorption lines of neutral
iron depend only on \teff, \vturb, and iron abundance, and are practically
independent of the surface gravity. On the other hand the lines of
ionised iron are sensitive to the temperature, metallicity and most of 
all to gravity. First, we adjust \vturb\ until there is no
correlation between iron abundances and line intensity for the Fe-I
lines. Second, \teff\ is adjusted until there is no trend in the
abundance versus excitation potential of the atomic level causing the
Fe-I lines. Both \vturb\ and \teff\ are
not independent, however the dependence on \vturb\
is stronger, so we adjust this parameter first. 
The gravity is obtained by fitting Fe-II and Fe-I lines and requiring
the same abundances from both neutral and ionised lines. 

\subsection{SOU}
The SOU parameters were derived starting with the automatic measurement of equivalent 
widths of Fe-I and Fe-II lines with
ARES \citep{sou07} and then imposing excitation and ionization 
equilibrium using a spectroscopic analysis in LTE
with the help of the code MOOG \citep{sne73} and a GRID of 
Kurucz Atlas 9 plane-parallel model atmospheres \citep{kur93}.
The Fe I and Fe II line list is composed of more than 300 lines that were 
individually tested in high resolution spectra
to check its stability to an automatic measurement with ARES. 
The atomic data of the lines were obtained from the Vienna
Atomic Line Database \citep{kup99} 
but the $\log gf$ were recomputed through an inverse analysis of the solar spectrum allowing
in this way to perform a differential analysis relative to the Sun. 
A full description of the method can be found in \citet{sou08}.
The errors on the SOU parameters are obtained by 
quadratically adding 60 K, 0.1 and 0.04 dex to the
internal errors on \teff, \logg, and [Fe/H], respectively. 
These values were obtained considering the typical dispersion plotted in
each comparison of parameters presented in \citet{sou08}. 
A more complete discussion about the errors derived for this
spectroscopic method can be seen in \citet{sou11}.

\subsection{VWA}
The VWA software \citep{bru10} relies on spectral synthesis of the region around individual Fe lines, but includes the contribution from weakly blending lines. The selection of lines was chosen from the degree of blending. 
These were then fitted with a synthetic spectrum. 
We adopted the MARCS model atmospheres \citep{gus08} and used the atomic line data from the Vienna Atomic Line Database \citep{kup99}. 
Each line fit was inspected in great detail and bad fits discarded, resulting in between 100--150 \feone/\fetwo\ lines that were used in the parameter determination. 
The oscillators strengths $\log gf$ were corrected relative to the Sun as done in \citet{bru10}.
The values quoted in the Kepler Input Catalogue were used as initial guesses for the parameters of the atmosphere. 
The parameters were then refined through several iterations where each parameter was adjusted so that any correlations between the abundances of \feone\ and equivalent width (EW) and excitation potential (EP) were removed. Also, we required agreement between the \feone\  and \fetwo\ abundances. Only lines with an ${\rm EW}\leq90\,m$\AA\ were used, since these lines are the most sensitive to changes in the fundamental parameters of the star.

To calculate the uncertainties of \teff, \logg, and $\xi_{\rm t}$ the model parameters were changed one at a time, until at least a 3$\sigma$ deviation was produced on the slope of the \feone-abundance vs.\ EP or EW, or between the \feone/\fetwo\ abundances. From these, the 1$\sigma$ error was calculated, giving the internal precision of the parameters calculated in VWA \citep{bru08}. The uncertainty on \feh\ was calculated as the standard error on the mean from Fe-I lines and adding quadratically the uncertainty from the model \teff, \logg, and $\xi_{\rm t}$.

To put further constraints on the \logg\ value of the targets stars, we also used individual fits of the pressure-sensitive Mg\,{\sc i}b lines and the Ca lines at $\lambda6122$\AA, $\lambda6162$\AA$\ $ and $\lambda6439$\AA. Each line was fitted with three different synthetic spectra, each with a different value for \logg. For each fit the $\chi^2$ was calculated and from this the best value of \logg\ was determined. The method is described in greater detail in a paper by \cite{bru10}. The final values for \logg\ were taken as the weighted mean of the values found from matching the \feone/\fetwo\ abundances and from fitting the pressure-sensitive lines.

\end{document}